\documentclass[fleqn,10pt]{wlpeerj}

\usepackage[utf8]{inputenc} 
\usepackage[T1]{fontenc}    
\usepackage{hyperref}       
\usepackage{url}            
\usepackage{booktabs}       
\usepackage{amsfonts}       
\usepackage{amsmath}
\usepackage{nicefrac}       
\usepackage{microtype}      
\usepackage{lipsum}
\usepackage{graphicx}
\graphicspath{ {./images/} }
\usepackage{svg}
\usepackage{float}
\usepackage{lineno}

\usepackage{subcaption}

\usepackage{amssymb}

\title{Quantitative Movement Testing: Measuring Chronic Pain Patient Movements from a Single Smartphone Video}

\author[1,2]{Pranav Mahajan}
\author[1]{Amanda Wall}
\author[1]{Eleonora Maria Camerone}
\author[3]{Julie Stebbins}
\author[1,4,5]{Eoin Kelleher}
\author[6]{Shuangyi Tong}
\author[1]{Annina Schmid}
\author[1]{Katja Wiech}
\author[1,7]{Anushka Irani}
\author[1,6]{Ben Seymour}
\affil[1]{Nuffield Department of Clinical Neurosciences, University of Oxford, Oxford, UK.}
\affil[2]{Max Planck Institute of Biological Cybernetics, Tuebingen, Germany.}
\affil[3]{Nuffield Department of Orthopaedics, Rheumatology and Musculoskeletal Sciences, University of Oxford, UK.}
\affil[4]{Harvard Medical School, Boston, Massachusetts, USA.}
\affil[5]{Massachusetts General Hospital, Boston, Massachusetts, US.}
\affil[6]{Institute of Biomedical Engineering, University of Oxford, Oxford, UK.}
\affil[7]{Mayo Clinic, Florida, USA.}

\corrauthor[]{Pranav Mahajan, Ben Seymour}{pranav.mahajan@ndcn.ox.ac.uk, ben.seymour@ndcn.ox.ac.uk}

\begin{abstract}
\textbf{BACKGROUND}
\newline
Chronic pain diminishes quality of life by decreasing functional ability, yet objectively measuring this functional impact, rather than relying solely on subjective pain reporting, remains challenging in real-world settings. While optical motion capture provides high precision for assessing altered movement quality, it is costly and restricted to laboratory environments. We aimed to develop and validate Quantitative Movement Testing (QMT), a computer vision pipeline extracting 3D kinematic biomarkers from standard monocular smartphone video, balancing clinical accessibility with biomechanical accuracy.
\newline
\textbf{METHODS}
\newline
We validated the QMT pipeline, utilising deep learning-based 3D pose-estimation, against gold-standard optical motion capture in healthy controls (N=13) across three tasks: forward bending, marching on the spot, and the minisquat. Following leave-one-subject-out calibration to correct systematic bias, we deployed QMT in two prospective clinical cohorts to assess real-world utility: a pre- and post-intervention trial for fibromyalgia patients (PainLESS; N=80 enrolled, N=57 QMT data captured), and a 30-day longitudinal at-home monitoring study of chronic sciatica patients and healthy controls (BeADS; N=97 enrolled, N=83 analysed).
\newline
\textbf{RESULTS}
\newline
In laboratory validation using healthy participants, QMT extracted clinical kinematic metrics with high agreement to optical motion capture, yielding strong correlations ($r>0.85$) and low mean absolute errors for most metrics. QMT outputs better matched planar (2D) surrogate angles (e.g., lumbar and knee flexion) than traditional Euler (3D) derivations (e.g., planar lumbar flexion $r=0.97$ vs. Euler $r=0.92$ during the minisquat). QMT demonstrated high test-retest reliability ($r > 0.86$) in fibromyalgia patients and successfully tracked day-to-day movement fluctuations in chronic sciatica. While real-world home settings introduced higher measurement variance than lab settings, QMT found group-level differences between healthy controls from sciatica patients based entirely on remote recordings.
\newline
\textbf{CONCLUSIONS}
\newline
By establishing a validated, open-source protocol and benchmarking it against current gold-standard optical motion capture, monocular 3D pose estimation offers a scalable alternative to traditional assessments. QMT provides an objective, accessible biomarker for tracking disease progression and treatment response in large-scale clinical trials, though further research is needed to optimise reliability in home-based environments.
\end{abstract}

\begin{document}

\flushbottom
\maketitle

\section*{Introduction}

Musculoskeletal conditions and chronic pain syndromes affect approximately 1.71 billion people globally, representing the leading contributor to disability worldwide \citep{cieza2020global}. 
For these patients, the decline in quality of life is driven by decrease in functional ability; a common feature of these conditions is altered movement quality, manifesting as reduced range of motion, motor variability, or protective guarding. Crucially, these functional changes are not always associated with mechanical tissue injury or proportional to the degree of any structural damage present. For instance, conditions driven by nociplastic or neuropathic mechanisms, such as fibromyalgia and sciatica, display distinct motor signatures often exacerbated by kinesiophobia (fear of movement) \citep{vlaeyen2000fear, vlaeyen2016fear} and central sensitisation \citep{woolf2011central, kosek2021chronic}. Yet, measuring these motor changes remains a challenge. Just as Quantitative Sensory Testing (QST), a standardised, non-invasive diagnostic method used to measure a person's sensory perception and pain thresholds, provides a semi-objective assessment of somatosensory function \citep{rolke2006quantitative}, the field lacks an equivalent "Quantitative Movement Testing" (QMT) to objectively profile a patient's functional capacity affordably and conveniently from their home. Currently, clinicians predominantly rely on subjective visual appraisal or static goniometry, neither of which captures the dynamic quality of movement required to monitor disease progression or treatment response in the real world.

Despite the urgency to establish these real-world biomarkers, the current gold standard, optical motion capture, remains tethered to the laboratory. While highly precise, these systems are prohibitively expensive (requiring dedicated laboratory infrastructure and substantial capital investment), demand hours of expert labour, and lack ecological validity, as patients may use altered movement strategies when observed in laboratory-based environments \citep{parsons2015virtual}. Further, for chronic pain populations with sensory hypersensitivity or tactile allodynia, the standard protocol of attaching reflective markers to the skin can be physically uncomfortable, rendering the method impractical and inequitable. Furthermore, the operational logistics of laboratory visits make systematic longitudinal tracking difficult in patient populations. Because functional impairment is sensitive to daily fluctuations, longitudinal monitoring has emerged as a core ingredient for digital medicine. As noted by \citet{kimmel2025human}, the hurdles to deploying movement as a biomarker are now transitioning from purely technological to operational, necessitating tools that can measure high-fidelity kinematics without the friction of optical motion capture at a gait lab.

Recent advances in computer vision have lowered these barriers, yet a translational gap remains. Multi-camera systems like OpenCap \citep{uhlrich2023opencap} have democratised research-grade biomechanics, proving valuable for applications like neuromuscular disease assessment \citep{ruth2025video}. However, such triangulation-based systems need two or more synchronised devices, creating logistical friction that limits independent home use. Conversely, single-camera approaches \citep{kidzinski2020deep} can successfully estimate clinical kinematic-based metrics from standard video. Yet, these approaches often rely on end-to-end deep learning architectures to map 2D keypoints \citep{cao2019openpose} directly to summary scores; this approach requires large training datasets (1000+ participants), has reduced geometric interpretability compared to 3D poses, and may not generalise beyond the use case it is trained for. Importantly, prior approaches have not been applied to measure the functional impact of chronic pain. These prior works focus predominantly on gait and not discrete functional tasks, such as lumbar flexion or mini-squats, where the movement signatures of chronic pain are most evident.

\begin{figure}[htbp]
    \centering
    \includegraphics[width=\textwidth]{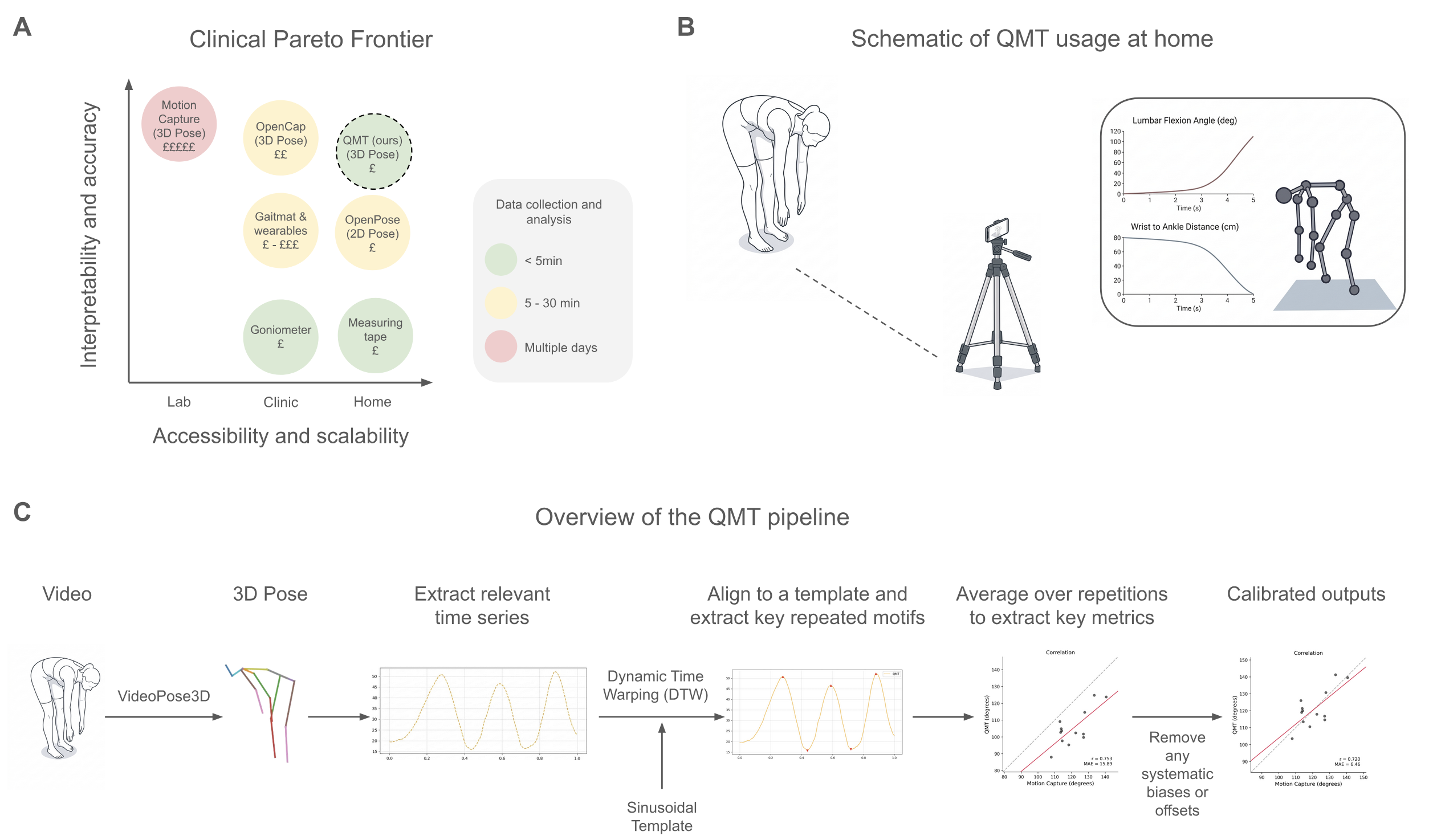} 
    \caption{\textbf{Overview of the Quantitative Movement Testing (QMT) framework.} (A) The clinical Pareto frontier illustrating the trade-off between biomechanical interpretability/accuracy and clinical accessibility/scalability and where QMT fits on the frontier. (B) Schematic demonstrating the at-home usage of QMT. A patient performs functional tasks (e.g., forward bend) in front of a single smartphone, from which 3D kinematic trajectories are extracted. (C) The QMT computational pipeline. Standard monocular video is processed via VideoPose3D to estimate 3D skeletal poses. Relevant kinematic time series are extracted, temporally aligned using Dynamic Time Warping (DTW) against a sinusoidal template, to segment the timeseries according to key predefined repeating motifs, which are then averaged across repetitions, and any systematic biases are removed to yield calibrated, clinically interpretable metrics.}
    \label{fig:Fig1}
\end{figure}

To address this gap, we present a bespoke Quantitative Movement Testing (QMT) framework designed specifically to navigate the "clinical Pareto frontier": intentionally balancing the accessibility of a single handheld (smartphone) camera with the interpretability, generalisability, and accuracy of 3D kinematic metrics (Fig. \ref{fig:Fig1}A). Unlike triangulation-based systems that require multiple synchronised devices \citep{uhlrich2023opencap}, our pipeline utilises a lifting architecture \citep{pavllo20193d} to infer 3D kinematics from monocular video (Fig. \ref{fig:Fig1}B). While this approach inherently trades distinct geometric certainty for improved accessibility, the clinical impact of this trade-off remains unbenchmarked for measurements of common pain-relevant movements. To evaluate its clinical utility, we devised a validation protocol co-designed with physiotherapists and pain patients to target specific movements. We benchmark the proposed system's accuracy against laboratory-based optical motion capture in healthy controls (N=13). We then demonstrate its clinical utility across the translational spectrum: first, in a randomised controlled trial of a sleep intervention for fibromyalgia (initial N=80), where we evaluate its sensitivity to treatment effects; and second, in a longitudinal 30-day "take-at-home" monitoring study of chronic sciatica patients and healthy controls (initial N=97), demonstrating the feasibility of capturing movement fluctuation in the real-world. Finally, to accelerate the adoption of objective movement profiling, we open-source the QMT software pipeline and protocol details.

\section*{Methods}

\subsection*{Participant Screening and Recruitment}

This study evaluated the QMT pipeline across three distinct datasets: a primary laboratory-based cohort for technical validation, and two prospective clinical cohorts to assess real-world utility.

\subsubsection*{Validation Dataset} 

Between February and May 2025, 13 healthy adults were recruited for laboratory validation. All participants provided written informed consent. The study received ethical approval from the University of Oxford Medical Sciences Interdivisional Research Ethics Committee (MS IDREC - 631993). 

Inclusion criteria were: aged 18 years or older, English-speaking, and independently mobile. Exclusion criteria were: any musculoskeletal or neurological condition affecting movement (e.g., low back, hip, knee, or ankle pain, or Parkinson's disease); musculoskeletal injury or surgery within the preceding six months; use of a joint brace or support; and current pregnancy.

\subsubsection*{Fibromyalgia Clinical Trial Dataset (PainLESS)} 

Between July 2023 and October 2024, 513 patients with fibromyalgia were approached, of whom 241 adult patients were assessed for eligibility. Of these, 80 met the inclusion criteria and were randomly assigned to the digital cognitive behavioural therapy for insomnia (dCBT-I) intervention Sleepio (n=42) or a control group (n=38). All participants provided written informed consent. The study was a registered clinical trial (NCT05962138) and received ethical approval from the South Central – Oxford B Research Ethics Committee (19/SC/0168).

Inclusion criteria were: aged 18 years or older; English speaking; a clinical diagnosis of fibromyalgia; concomitant insomnia (characterised by frequent night-time or early morning waking); self-reported cognitive difficulties; reliable internet access; and ability to read and understand English. Exclusion criteria were: known neurological conditions; major neuropsychiatric disorders (excluding depression and anxiety); epilepsy; cognitive impairment or neurodegenerative disorders; recent or planned surgery; current or planned night-shift work; specific untreated sleep disorders (e.g., sleep apnoea, restless leg syndrome); use of prescribed sleep medications on more than two nights in the past two weeks; currently receiving other psychological therapy for insomnia; and pregnancy or lactation. 

Of the 80 participants with fibromyalgia who attended a pre-intervention visit, 63 attended a post-intervention visit, and of these, a further 9 declined to perform some or all of the video recordings on the day of their research visit. Thus, the final analytical sample comprised 54 subjects with paired pre- and post-intervention forward bend recordings, and 57 subjects with paired pre- and post-intervention marching on the spot and minisquat recordings.

\subsubsection*{Longitudinal Sciatica Dataset (BeADS)}

A total of 46 healthy individuals and 51 patients with chronic sciatica were enrolled in the 30-day longitudinal study. Following 6 withdrawals from the patient cohort, 45 patients completed the study. Prior to participation, all individuals provided written informed consent to a protocol approved by the Oxford Research Ethics Committee (18/SC/0263).

Inclusion criteria for chronic sciatica participants were: age 16–75 years; sciatica symptoms present for $\geq$ 6 months, with no complete remission during this period (defined as pain on $\geq$ 50\% of days in the past 6 months; with no symptom-free interval exceeding 3 months). Exclusion criteria were: serious spinal pathology (e.g., lumbar spine surgery); severe psychological or psychiatric condition; diagnosed diabetic neuropathy, or diabetes with bilateral foot pain, pins and needles, or tingling (diabetes alone without these symptoms was not exclusionary); known neurological conditions; systemic inflammatory disease; moderate-to-severe pain from other conditions that may confound pain assessment; current or scheduled cancer treatment; insufficient mental capacity or insufficient command of English to consent to the study or complete questionnaires; deemed unsuitable for participation by the investigator; and pregnancy.
Inclusion criteria for healthy participants were: age 16–75 years. Exclusion criteria were: history of any peripheral or central neuropathy; current or past chronic pain; current acute pain in the lower body (back/legs); and all further exclusion criteria as applied to the chronic sciatica cohort.

Out of 91 subjects who completed the study (46 healthy, 45 patients), 2 patients failed to upload videos due to technical difficulties from the second day onward. Following manual quality-control exclusions (e.g., framing errors such as completely truncated heads or ankles, portrait-mode recording, or failure to capture the functional movement), 209 of the 2,539 submitted videos (i.e. 8.23\%) were manually excluded.
This resulted in a final dataset of 44 healthy controls and 39 sciatica patients (after manual exclusions) whose data was used in QMT validation.

\subsection*{Study Procedures and Protocol Design}

The clinical protocols were built around four functional movement categories: forward bend, minisquat, marching on the spot and heel raise. These movements were specifically selected and standardised into a reproducible protocol because they represent actions that are used in day-to-day activities such as bending to pick something up, squatting to sit down, going up stairs or a step. They are also movements that highlight functional deficits commonly observed in chronic pain conditions, including chronic low back pain, sciatica, fibromyalgia, and osteoarthritis.  

\subsubsection*{Laboratory Validation Protocol} 
Participants in the healthy validation cohort wore tight, sports-type clothing and were barefoot. Passive reflective markers were applied to anatomical landmarks according to a modified conventional gait model \citep{davis1988clinical}, with additional task-specific markers in line with Oxford Gait Laboratory protocols. Marker identities and placement descriptions are provided in Supplementary Table S1. Participants were instructed to move in their typical manner, to an extent, and at a pace, that they felt comfortable with, rather than to their maximal possible range or as fast as they could go. Kinematic gait laboratory data were simultaneously recorded using a 16-camera Valkyrie motion capture system (Vicon, Oxford, UK) recording at 100 Hz, plus a 4-camera Flir video system (Vicon, Oxford, UK), recording at 100Hz (sagittal and coronal views, full body and lower limb close up views); 3 force plates (AMTI) were also used, recording at 100Hz. The gait laboratory software was Nexus version 2.15 (Vicon, Oxford UK).

\subsubsection*{PainLESS Trial Protocol} 
The fibromyalgia cohort followed a pre- and post-intervention workflow. The researcher recorded participants performing up to three repetitions of the forward bend, marching on the spot and minisquat at baseline, in a standardised research setting, and then repeated the identical recording protocol following the completion of the dCBT-I intervention (or control period). 

\subsubsection*{BeADS Longitudinal Protocol}

Both cohorts  completed a 30-day unsupervised at-home monitoring protocol focusing exclusively on lumbar spine forward bending, following a single supervised practice trial conducted with the experimenter in the experimental setting to ensure correct task understanding. Each day, participants performed a forward-bending movement to their maximum comfortable range. They were instructed to complete the task preferably in the morning; if this was not possible, they were asked to select a consistent time of day and adhere to it throughout the protocol. Participants were provided with a standardised kit comprising a tripod, a metric tape measure, adhesive fixings (blu-tack or velcro), and a daily rating diary.

Daily forward-bend movements were video-recorded using the provided tripod and subsequently analysed using the QMT pipeline. To ensure positional consistency across days, participants received detailed instructions on tripod placement relative to their standing position (45° angle), and the tripod location was marked with adhesive fixings to minimise day-to-day variability in setup (Fig. \ref{fig:Fig1}B). Forward bending range was additionally quantified via self-report using a wall-mounted metric tape, which allowed participants to estimate their fingertip-to-floor distance.

Each day, participants completed ratings before and after the movement using their daily rating diary, submitted via REDCap. Prior to bending, they rated their mood (1–5 scale), expected movement ability (fingertip-to-floor distance), and anticipated pain intensity (Numeric Rating Scale: 0 = no pain, 10 = unbearable pain). During the forward bend, participants visually noted their actual reach (minimum fingertip-to-floor distance) using the wall-mounted tape. Following the movement, they reported their actual reach distance, pain experienced during the task, and predicted functional capacity and pain for the following day.

\subsection*{Kinematic Metric Derivation}

The QMT pipeline derived functional metrics from a 17-joint VideoPose3D skeleton \citep{pavllo20193d} comprising the pelvis, bilateral hips, knees, ankles, spine, upper torso, neck base, head, shoulders, elbows, and wrists. From these predicted 3D joint trajectories, we computed movement-specific planar (2D) angles and spatial distances to maximise geometric stability. 

Lumbar flexion was defined as the angle of the spine-pelvis vector (as derived from the VideoPose3D skeleton) relative to a fixed vertical reference axis. Knee flexion was calculated as the angle between the hip-knee and knee-ankle vectors, derived from the VideoPose3D skeleton. Hip flexion was defined as the angle between the spine-pelvis and hip-knee vectors, while trunk sway was the angle between the spine-pelvis and hip-pelvis vectors in the frontal plane, derived from VideoPose3D skeleton. Spatial functional metrics included reach during the forward bend (the vertical distance between the mean wrist position and the mean ankle position, at the maximum bend) and pelvis depth during the minisquat (the vertical distance between the pelvis node and the mean ankle position, at maximum squat depth).

To validate these QMT derivations against the gold standard optical motion capture, we paired each QMT metric with its closest available biomechanical match (the equivalent planar angle). Because the lower-dimensional pose-estimation skeleton does not retain the full marker-set geometry available to motion capture, this required calculating both traditional 3D Euler angles and planar (2D) surrogate angles from the C3D motion capture files. Primary reference targets included trunk Euler angles, functional knee and hip Euler angles, and spatial marker distances (e.g., wrist-to-ankle reach  \citep{stebbins2025natural}). To generate planar (2D) surrogate angles that anatomically matched the QMT representations, we mapped specific motion capture markers to the VideoPose3D nodes; for example, after comparing the gait laboratory T2, T6, and T12 markers, the T6 marker was selected as it best anatomically mapped to the QMT `spine' node for calculating planar lumbar flexion and sway.












\subsection*{The Quantitative Movement Testing (QMT) analysis pipeline}

The QMT pipeline is summarised in Fig. \ref{fig:Fig1}C. The QMT pipeline estimated framewise 2D and 3D pose trajectories from each video using a pre-trained VideoPose3D model \citep{pavllo20193d}. Videos were primarily recorded in landscape format to optimise VideoPose3D performance. The resulting skeletal coordinates were used to derive the specific planar and spatial kinematic variables detailed above. 

To enable comparison across recordings of different durations, each trajectory was interpolated onto a normalised time axis. Repetition-level analysis was performed using Dynamic Time Warping (DTW) \citep{giorgino2009computing} against a cosine template, employing a constrained alignment to identify the temporal structure of each movement cycle. Template extrema (peak and troughs) were mapped back to the observed signal to yield repetition-wise minima (trough), maxima (peak), and derived functional ranges (peak minus trough). Downstream clinical metrics were derived by averaging extrema over all repetitions.

\subsection*{Statistical Analysis}

Statistical analyses were structured around the three distinct study branches: validation, therapeutic intervention (PainLESS), and longitudinal monitoring (BeADS).

\subsubsection*{Laboratory Validation} 
Agreement between QMT and optical motion capture was evaluated using Pearson correlation ($r$) mean absolute error (MAE), systematic bias, and Bland-Altman limits of agreement \citep{giavarina2015understanding}. To correct for systemic algorithmic bias inherent in monocular depth estimation, a leave-one-subject-out cross-validation (LOOCV) calibration was applied. For each held-out subject, an additive offset was estimated from the remaining subjects as the mean difference between motion-capture and QMT values. This offset was then applied to the held-out subject's measurements. To assess robustness for patient-administered use, QMT kinematics were summarised and compared using the mean of 10 trials, 3 trials, and a single trial.  Lastly, across all results, we use Pearson correlations, similar to \citep{uhlrich2023opencap, ruth2025video} instead of intraclass correlations (ICC) due to low $N=13$, as stable ICC estimates often require $N \geq 30$.

\subsubsection*{PainLESS Trial} 
Analyses were performed using paired pre- and post-intervention QMT summaries. The primary analysis compared subject-level change scores (post-intervention minus pre-intervention) between the dCBT-I and control groups using Welch’s two-sample t-tests to account for unequal variances. For each comparison, Hedges’ $g$ was calculated to estimate effect size, alongside a Bayes Factor ($BF_{10}$) to quantify the weight of evidence for or against a between-group difference. Secondary analyses included within-group paired t-tests and subject-level mixed-effects models with fixed effects for time, group, and their interaction, and a random intercept for the participant. False discovery rate (FDR) correction was applied across metrics to account for multiple comparisons.

\subsubsection*{BeADS Longitudinal Study}
To attempt a direct quantitative comparison between QMT reach distance (wrist-to-ankle) and patient self-reports (fingertip-to-floor), QMT values were aligned to the self-report scale using a LOOCV additive calibration offset ($-16.46$ cm globally). Agreement was summarised using subject-level MAE, Pearson correlation, within-subject standard deviation, and mean absolute error in day-to-day changes in QMT outcomes and self-reports ($\Delta$ MAE). Finally, functional differences between the independent clinical cohorts (Healthy vs. Sciatica) were assessed using Welch's two-sample t-tests.

\section*{Results}

\subsection*{Validation of QMT against lab-based gold standard optical motion capture}

\begin{figure}[htbp]
    \centering
    \includegraphics[width=\textwidth]{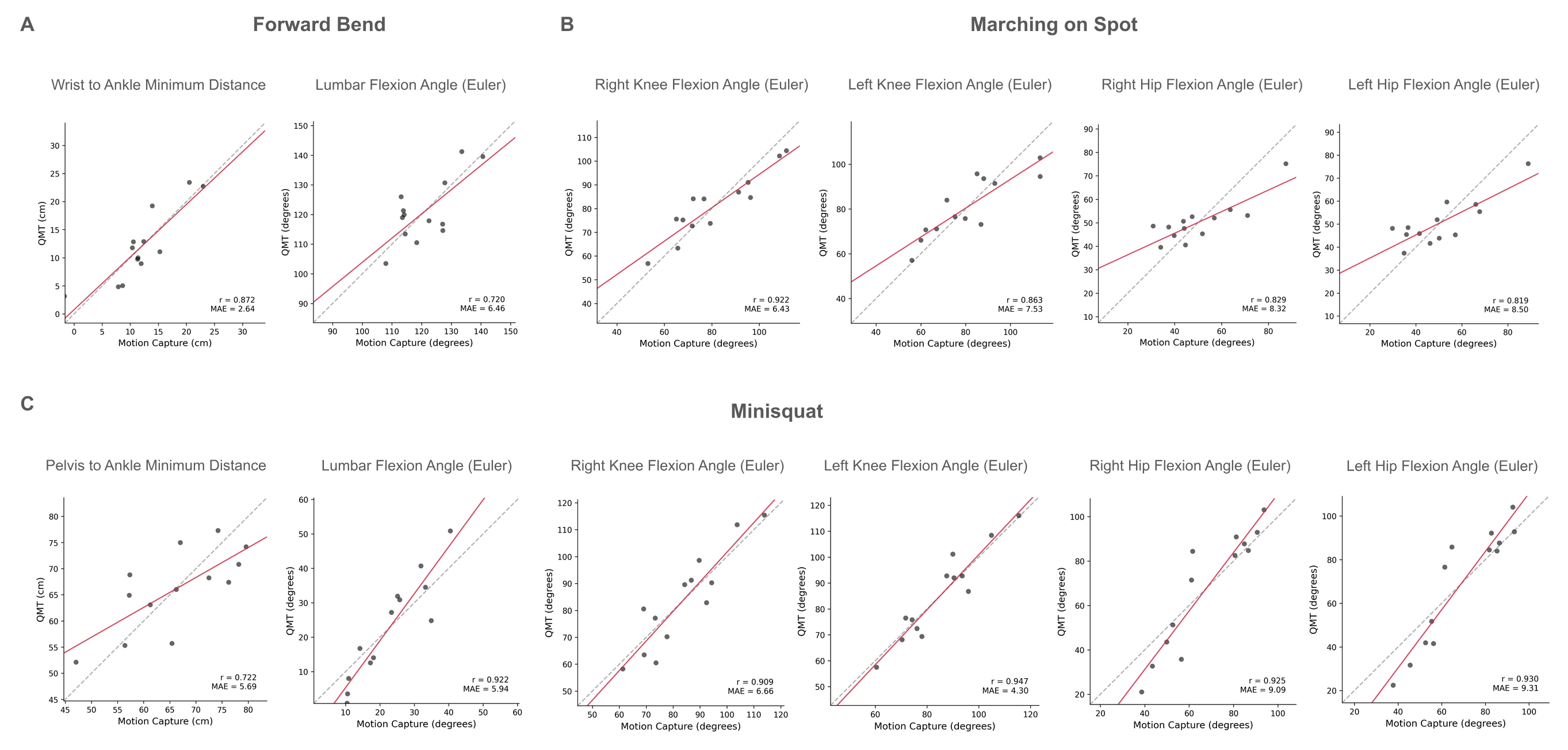} 
    \caption{\textbf{Primary Outcomes.} Comparison of Quantitative Movement Testing (QMT) and gold-standard optical motion capture for spatial distances and traditional 3D Euler angles The Pearson correlations (r) and Mean Absolute Errors (MAEs) demonstrate strong agreement across functional tasks following leave-one-subject-out calibration.}
    \label{fig:primary_outcomes}
\end{figure}

\begin{figure}[htbp]
    \centering
    \includegraphics[width=0.8\textwidth]{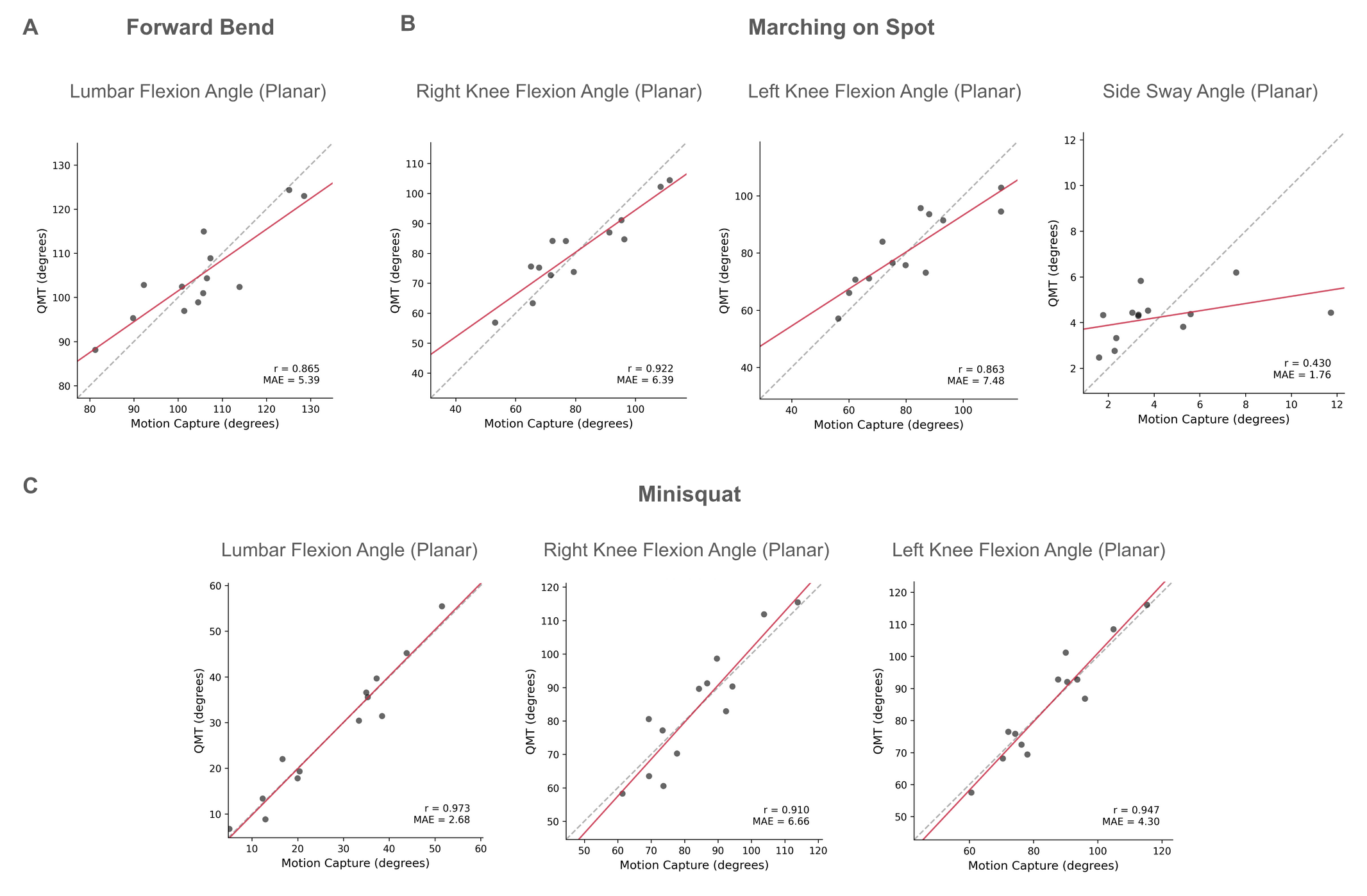} 
    \caption{\textbf{Planar (2D) Surrogate Angles.} Validation of planar angle projections designed to mitigate monocular depth ambiguities. Planar derivations demonstrated improved reliability for trunk motion kinematics (e.g., lumbar flexion) compared to traditional 3D Euler derivations, while maintaining high accuracy for appendicular joints (except, sway angle).}
    \label{fig:planar_surrogates}
\end{figure}

To evaluate the concurrent validity of the Quantitative Movement Testing (QMT) pipeline, we compared its kinematic outputs to concurrent 3D optical motion capture in healthy controls (N=13) across three functional tasks: forward bending, marching on the spot, and the minisquat. The validation analysis was stratified into two domains: an evaluation of primary clinical outcomes, comprising spatial distances and traditional 3D Euler angles from motion capture (Figure \ref{fig:primary_outcomes}), followed by an additional comparison with planar surrogate angles which best matched the 2D planar angles provided by QMT (Figure \ref{fig:planar_surrogates}). 

We report both uncalibrated and LOOCV-calibrated performance in Table~\ref{tab:calibration_summary}, treating calibration as a bias-correction step. Prior to calibration, raw QMT measurements demonstrated strong linear relationships with motion capture but exhibited varied systematic biases. Some measurements, such as wrist-to-ankle minimum distance, pelvis-to-ankle minimum distance, and planar lumbar flexion angle, had negligible bias. Conversely, all knee and hip flexion measurements, as well as Euler lumbar flexion angles, exhibited significant systematic bias (Table \ref{tab:calibration_summary}). Following a leave-one-subject-out cross-validation (LOOCV) calibration process, systematic bias was eliminated (Bias = 0.00) across all evaluated metrics (Table \ref{tab:calibration_summary}). 
Participant-specific LOOCV offsets are provided in Supplementary Table S2; for most movement-metric pairs these applied offsets fell within relatively narrow ranges, although this does not entirely eliminate the concerns of generalisability associated with the small validation sample.
Note that while heel raises were recorded, they were subsequently excluded from the final analyses and results, as changes in absolute ankle height could not be reliably estimated from the relative root-node coordinates of the 3D pose skeleton.

\begin{table}[htbp]
\centering
\caption[Summary of Validation Results Across Three Movements before and after calibration (bias correction)]{\textbf{Summary of Validation Results Across Three Movements before and after calibration (bias correction)}. LoA denotes Bland-Altman Limits of Agreement. Comparison is calculated as QMT measurement minus motion capture measurement. Thus negative raw bias means the average QMT measurement was lower than the average measurement from motion capture, and vice versa. Note that the bias (Cal Bias) is 0 for all metrics after calibration. Results after calibration (Cal LoA, Cal r and Cal MAE) are calculated using leave-one out cross validation (LOOCV). The correlation plots and Bland-Altman plots for all of the following metrics before and after calibration, with more details are provided in Supplementary Figures S1-S5.}
\label{tab:calibration_summary}
\resizebox{\textwidth}{!}{
\begin{tabular}{llrrrrrrrrrr}
\toprule
\textbf{Movement} & \textbf{Metric} & \textbf{N} & \textbf{Raw Bias} & \textbf{Raw LoA Lower} & \textbf{Raw LoA Upper} & \textbf{Raw r} & \textbf{Raw MAE} & \textbf{Cal LoA Lower} & \textbf{Cal LoA Upper} & \textbf{Cal r} & \textbf{Cal MAE} \\
\midrule
Forward Bend & Wrist to Ankle Min Dist & 13 & -0.53 & -6.04 & 4.99 & 0.89 & 2.49 & -5.98 & 5.98 & 0.87 & 2.64 \\
Forward Bend & Lumbar Flexion (Euler) & 13 & -15.89 & -29.36 & -2.41 & 0.75 & 15.89 & -14.60 & 14.60 & 0.72 & 6.46 \\
Forward Bend & Lumbar Flexion (Planar) & 13 & 0.35 & -11.10 & 11.79 & 0.89 & 4.95 & -12.40 & 12.40 & 0.86 & 5.39 \\
\midrule
Marching on Spot & Right Knee Flexion (Euler) & 13 & -23.21 & -36.26 & -10.15 & 0.94 & 23.21 & -14.14 & 14.14 & 0.92 & 6.43 \\
Marching on Spot & Right Knee Flexion (Planar) & 13 & -23.22 & -36.19 & -10.26 & 0.94 & 23.22 & -14.05 & 14.05 & 0.92 & 6.39 \\
Marching on Spot & Left Knee Flexion (Euler) & 13 & -26.53 & -43.12 & -9.95 & 0.89 & 26.53 & -17.97 & 17.97 & 0.86 & 7.53 \\
Marching on Spot & Left Knee Flexion (Planar) & 13 & -26.57 & -43.08 & -10.05 & 0.89 & 26.57 & -17.89 & 17.89 & 0.86 & 7.48 \\
Marching on Spot & Right Hip Flexion (Euler) & 13 & -31.29 & -48.63 & -13.96 & 0.87 & 31.29 & -18.78 & 18.78 & 0.83 & 8.32 \\
Marching on Spot & Left Hip Flexion (Euler) & 13 & -30.09 & -47.53 & -12.65 & 0.86 & 30.09 & -18.90 & 18.90 & 0.82 & 8.50 \\
Marching on Spot & Side Sway (Planar) & 13 & 0.51 & -3.91 & 4.94 & 0.59 & 1.82 & -4.79 & 4.79 & 0.43 & 1.76 \\
\midrule
Minisquat & Pelvis to Ankle Min Dist & 13 & -0.91 & -12.88 & 11.05 & 0.77 & 5.33 & -12.96 & 12.96 & 0.72 & 5.69 \\
Minisquat & Lumbar Flexion (Euler) & 13 & 7.56 & -4.45 & 19.57 & 0.93 & 8.05 & -13.01 & 13.01 & 0.92 & 5.94 \\
Minisquat & Lumbar Flexion (Planar) & 13 & 2.48 & -3.43 & 8.39 & 0.98 & 3.35 & -6.40 & 6.40 & 0.97 & 2.68 \\
Minisquat & Right Knee Flexion (Euler) & 13 & -11.56 & -25.03 & 1.91 & 0.92 & 11.56 & -14.59 & 14.59 & 0.91 & 6.66 \\
Minisquat & Right Knee Flexion (Planar) & 13 & -11.60 & -25.04 & 1.84 & 0.92 & 11.60 & -14.56 & 14.56 & 0.91 & 6.66 \\
Minisquat & Left Knee Flexion (Euler) & 13 & -12.69 & -22.46 & -2.92 & 0.95 & 12.69 & -10.59 & 10.59 & 0.95 & 4.30 \\
Minisquat & Left Knee Flexion (Planar) & 13 & -12.72 & -22.47 & -2.97 & 0.95 & 12.72 & -10.56 & 10.56 & 0.95 & 4.30 \\
Minisquat & Right Hip Flexion (Euler) & 13 & -6.15 & -27.02 & 14.72 & 0.93 & 9.76 & -22.61 & 22.61 & 0.93 & 9.09 \\
Minisquat & Left Hip Flexion (Euler) & 13 & -8.42 & -28.95 & 12.10 & 0.94 & 11.31 & -22.23 & 22.23 & 0.93 & 9.31 \\
\bottomrule
\end{tabular}
}
\end{table}

\subsubsection*{Primary Outcomes: Spatial Distances and 3D Euler Angles}

For primary spatial metrics, the QMT pipeline accurately captured functional range of motion. During the forward bend, the functional reach metric (wrist-to-ankle minimum distance) demonstrated excellent agreement with motion capture post-calibration ($r = 0.87$, Mean Absolute Error [MAE] = 2.64 cm). Similarly, the depth of the minisquat (pelvis-to-ankle minimum distance) showed moderate-to-strong agreement ($r = 0.72$, MAE = 5.69 cm).

\subsubsection*{Additional Surrogate Comparisons: 2D Planar Angles}

For appendicular joints, planar derivations of knee flexion performed comparably to 3D Euler angles across both the marching and minisquat tasks. They maintained high correlations ($r$ ranging from $0.86$ to $0.95$) and low absolute errors (MAE $4.3^\circ$--$7.5^\circ$), indicating that planar (2D) surrogates are an equally valid alternative for single-plane hinge joints. We note that all videos were taken from the forward-right direction, which likely explains why left knee angle correlations were slightly lower than their respective right knee angle correlations. Finally, physiological trunk control was assessed via planar side sway during the marching task. While the correlation coefficient was lower ($r = 0.43$), the absolute error was clinically negligible (MAE = 1.76$^\circ$). This reduced correlation is largely an artefact of the highly restricted physiological range of motion for healthy sway.

\subsubsection*{Robustness to Trial Repetitions and Supplementary Analyses}
To assess the feasibility of reducing physical burden for patients during in-lab visits and remote at-home monitoring, we evaluated the pipeline's robustness across varying trial repetitions. Agreement metrics remained broadly similar when summarising 10 trials, 3 trials, or a single trial, with only modest changes in MAE and LoA for most measures (Supplementary Table S3). Because 10-trial laboratory averages are rarely practical in clinical populations, this single-trial stability is a critical prerequisite for the subsequent clinical deployments detailed in the following sections. Additionally, comprehensive subject-level correlation and Bland-Altman plots are provided in Supplementary Figures S1–S5, and participant-specific applied calibration offsets are reported in Supplementary Table S2, to ensure full methodological transparency.

\subsection*{Clinical utility of QMT in the PainLESS fibromyalgia trial}

We next evaluated the real-world clinical utility of the QMT pipeline by deploying it within the PainLESS trial, a randomised controlled study of patients with fibromyalgia. The objective was to determine whether monocular 3D pose estimation could successfully operate as a robust, longitudinal digital endpoint in a patient population characterised by pervasive functional deficits.

\subsubsection*{Test-retest reliability in a patient population}

Before evaluating longitudinal treatment effects, it is necessary to establish the baseline stability of the proposed digital endpoints. We assessed the test-retest reliability of the QMT pipeline in a subset of fibromyalgia patients (N=12) who provided two measurements of the full suite of functional tasks on the same day.

QMT demonstrated consistent intra-subject reliability across all evaluated metrics (Table \ref{tab:test_retest}). The planar (2D) surrogate angles proved stable, yielding strong correlations ranging from $r = 0.78$ to $0.99$ (all $p \le 0.003$). Notably, trunk motion metrics such as planar lumbar flexion during the minisquat ($r = 0.99$, ICC = $0.99$) and planar side sway during marching on the spot ($r = 0.99$, ICC = $0.96$) exhibited high agreement between sessions. In addition to Pearson correlation and ICC, we also provide MAE and Bland-Altman limits of agreement (LoA). However, we note that ICC can be unstable due low $N=12$, therefore we also provide the 95\% confidence intervals. This test-retest reliability indicates that the QMT pipeline can provide a stable quantitative measurement capable of detecting subtle kinematic changes in a chronic pain population.

\begin{table}[htbp]
\centering
\caption{\textbf{Test-Retest Reliability of QMT Metrics in Patients with Fibromyalgia.} Diff denotes Retest measurement minus Test measurement.}
\label{tab:test_retest}
\resizebox{\textwidth}{!}{
\begin{tabular}{llrrrrrrrrr}
\toprule
\textbf{Movement} & \textbf{Metric} & \textbf{N} & \textbf{r} & \textbf{p-value} & \textbf{ICC (95\% CI)} & \textbf{MAE} & \textbf{Mean Diff} & \textbf{SD Diff} & \textbf{LoA Lower} & \textbf{LoA Upper} \\
\midrule
Forward Bend & Wrist to Ankle Min Dist & 12 & 0.93 & <0.001 & 0.93 [0.79, 0.98] & 4.93 & -2.21 & 6.57 & -15.10 & 10.67 \\
 & Lumbar Flexion (Planar) & 12 & 0.92 & <0.001 & 0.92 [0.74, 0.98] & 4.84 & 2.11 & 5.95 & -9.54 & 13.76 \\
\midrule
Marching on Spot & Right Knee Flexion (Planar) & 12 & 0.86 & <0.001 & 0.85 [0.56, 0.95] & 7.34 & 2.52 & 8.92 & -14.97 & 20.01 \\
 & Left Knee Flexion (Planar) & 12 & 0.81 & 0.001 & 0.81 [0.46, 0.94] & 10.09 & -0.27 & 12.06 & -23.91 & 23.37 \\
 & Right Hip Flexion (Planar) & 12 & 0.87 & <0.001 & 0.85 [0.57, 0.96] & 10.57 & 2.48 & 12.75 & -22.51 & 27.48 \\
 & Left Hip Flexion (Planar) & 12 & 0.86 & <0.001 & 0.86 [0.59, 0.96] & 10.13 & 3.24 & 12.51 & -21.27 & 27.75 \\
 & Side Sway (Planar) & 12 & 0.99 & <0.001 & 0.96 [0.88, 0.99] & 2.05 & 1.45 & 3.69 & -5.79 & 8.70 \\
\midrule
Minisquat & Pelvis to Ankle Min Dist & 12 & 0.86 & <0.001 & 0.83 [0.51, 0.95] & 5.73 & 1.74 & 7.59 & -13.14 & 16.62 \\
 & Lumbar Flexion (Planar) & 12 & 0.99 & <0.001 & 0.99 [0.96, 1.00] & 3.45 & -1.39 & 4.45 & -10.13 & 7.34 \\
 & Right Knee Flexion (Planar) & 12 & 0.81 & 0.001 & 0.81 [0.46, 0.94] & 10.06 & 1.46 & 12.65 & -23.34 & 26.25 \\
 & Left Knee Flexion (Planar) & 12 & 0.78 & 0.003 & 0.77 [0.39, 0.93] & 9.50 & 0.90 & 12.06 & -22.73 & 24.54 \\
 & Right Hip Flexion (Planar) & 12 & 0.92 & <0.001 & 0.92 [0.74, 0.98] & 10.72 & 1.22 & 12.62 & -23.51 & 25.95 \\
 & Left Hip Flexion (Planar) & 12 & 0.94 & <0.001 & 0.94 [0.82, 0.98] & 8.67 & 2.47 & 10.70 & -18.51 & 23.45 \\
\bottomrule
\end{tabular}
}
\end{table}

\subsubsection*{QMT measurements as viable digital endpoints in clinical trials}

Having demonstrated test-retest reliability through high correlations, low mean absolute errors and bland altman limits of agreement, we next utilised the QMT pipeline to evaluate secondary functional outcomes in the PainLESS trial. We compared pre- and post-intervention kinematic change scores ($\Delta$) between patients randomised to dCBT-I and a control group. The final analytical sample sizes for these comparisons were N=54 for forward bending and N=57 for the minisquat and marching on the spot.

The primary Welch's two-sample t-test analysis revealed no statistically significant between-group differences in kinematic change scores across any of the functional movements (all $p > 0.05$). To determine whether this lack of significance was due to a true absence of clinical effect or simply insufficient statistical power, we computed Bayes Factors ($BF_{10}$) for all primary comparisons (Table \ref{tab:primary_painless}). The Bayes Factors consistently ranged between 0.14 and 0.64, providing moderate to anecdotal evidence in favour of the null hypothesis ($H_0$). Secondary within-group paired analyses and longitudinal mixed-effects models corroborated these findings, confirming the absence of a significant Group $\times$ Time interaction across all assessed metrics (Supplementary Tables S4 and S5).

\begin{table}[htbp]
\centering
\caption{Primary Group Comparison of Pre- to Post-Intervention Changes (PainLESS Trial). N (dCBT-I / Ctrl) refers to the number of participants in the intervention and control groups, respectively. $\Delta$ denotes the mean change from baseline to post-intervention (post-intervention minus pre-intervention). Evidence strength is based on the Bayes Factor ($BF_{10}$). }
\label{tab:primary_painless}
\resizebox{\textwidth}{!}{
\begin{tabular}{llrrrrrrrrr}
\toprule
\textbf{Movement} & \textbf{Metric} & \textbf{N (dCBT-I / Ctrl)} & \textbf{$\Delta$ dCBT-I} & \textbf{$\Delta$ Ctrl} & \textbf{Diff ($\Delta$)} & \textbf{Welch's t} & \textbf{p-value} & \textbf{Hedges' $g$} & \textbf{BF$_{10}$} & \textbf{Evidence} \\
\midrule
Forward Bend & Wrist to Ankle Min Dist (cm) & 28 / 26 & -2.77 & -4.48 & 1.71 & 0.35 & 0.727 & 0.10 & 0.15 & Moderate $H_0$ \\
~ & Lumbar Flexion (Planar) & 28 / 26 & 3.67 & 3.60 & 0.06 & 0.01 & 0.989 & 0.00 & 0.14 & Moderate $H_0$ \\
\midrule
Marching on Spot & Right Knee Flexion (Planar) & 30 / 27 & 2.11 & -0.44 & 2.55 & 0.57 & 0.572 & 0.15 & 0.16 & Moderate $H_0$ \\
~ & Left Knee Flexion (Planar) & 30 / 27 & -0.83 & -4.95 & 4.13 & 0.90 & 0.372 & 0.24 & 0.20 & Moderate $H_0$ \\
~ & Right Hip Flexion (Planar) & 30 / 27 & 7.00 & -0.55 & 7.55 & 1.36 & 0.179 & 0.35 & 0.33 & Moderate $H_0$ \\
~ & Left Hip Flexion (Planar) & 30 / 27 & 4.35 & -3.61 & 7.96 & 1.57 & 0.122 & 0.41 & 0.45 & Anecdotal $H_0$ \\
~ & Side Sway (Planar) & 30 / 27 & 4.37 & 0.08 & 4.30 & 1.81 & 0.080 & 0.45 & 0.59 & Anecdotal $H_0$ \\
\midrule
Minisquat & Pelvis to Ankle Min Dist (cm) & 30 / 27 & -3.95 & -1.47 & -2.47 & -0.50 & 0.622 & -0.13 & 0.15 & Moderate $H_0$ \\
~ & Lumbar Flexion (Planar) & 30 / 27 & 9.18 & 2.94 & 6.24 & 1.05 & 0.301 & 0.26 & 0.22 & Moderate $H_0$ \\
~ & Right Knee Flexion (Planar) & 30 / 27 & 3.49 & 3.24 & 0.25 & 0.04 & 0.969 & 0.01 & 0.13 & Moderate $H_0$ \\
~ & Left Knee Flexion (Planar) & 30 / 27 & 2.40 & 0.42 & 1.97 & 0.36 & 0.723 & 0.09 & 0.14 & Moderate $H_0$ \\
~ & Right Hip Flexion (Planar) & 30 / 27 & 8.84 & -0.99 & 9.83 & 1.48 & 0.146 & 0.39 & 0.40 & Anecdotal $H_0$ \\
~ & Left Hip Flexion (Planar) & 30 / 27 & 9.97 & -0.99 & 10.96 & 1.70 & 0.096 & 0.44 & 0.56 & Anecdotal $H_0$ \\
\bottomrule
\end{tabular}
}
\end{table}

Collectively, these results suggest the operational feasibility of the QMT pipeline for prospective clinical deployment. The absence of detected group differences is consistent with either a true null effect of dCBT-I on movement or with insufficient sensitivity of QMT to intervention-induced kinematic change. Without a positive control demonstrating responsiveness to known kinematic perturbations, these explanations cannot be disambiguated, and is an area for future work. The PainLESS data therefore establish feasibility of remote kinematic capture in a clinical trial workflow, but responsiveness validation requires future work.

\subsection*{Clinical feasibility of QMT in the longitudinal BeADS at-home study}

To determine the feasibility of QMT for continuous, remote patient monitoring, we deployed the pipeline in a 30-day home monitoring  study (BeADS). The cohort comprised patients with chronic sciatica and healthy controls who recorded daily forward bending tasks using their smartphones in an unconstrained home environment.

\begin{figure}[h!]
    \centering
    \includegraphics[width=\textwidth]{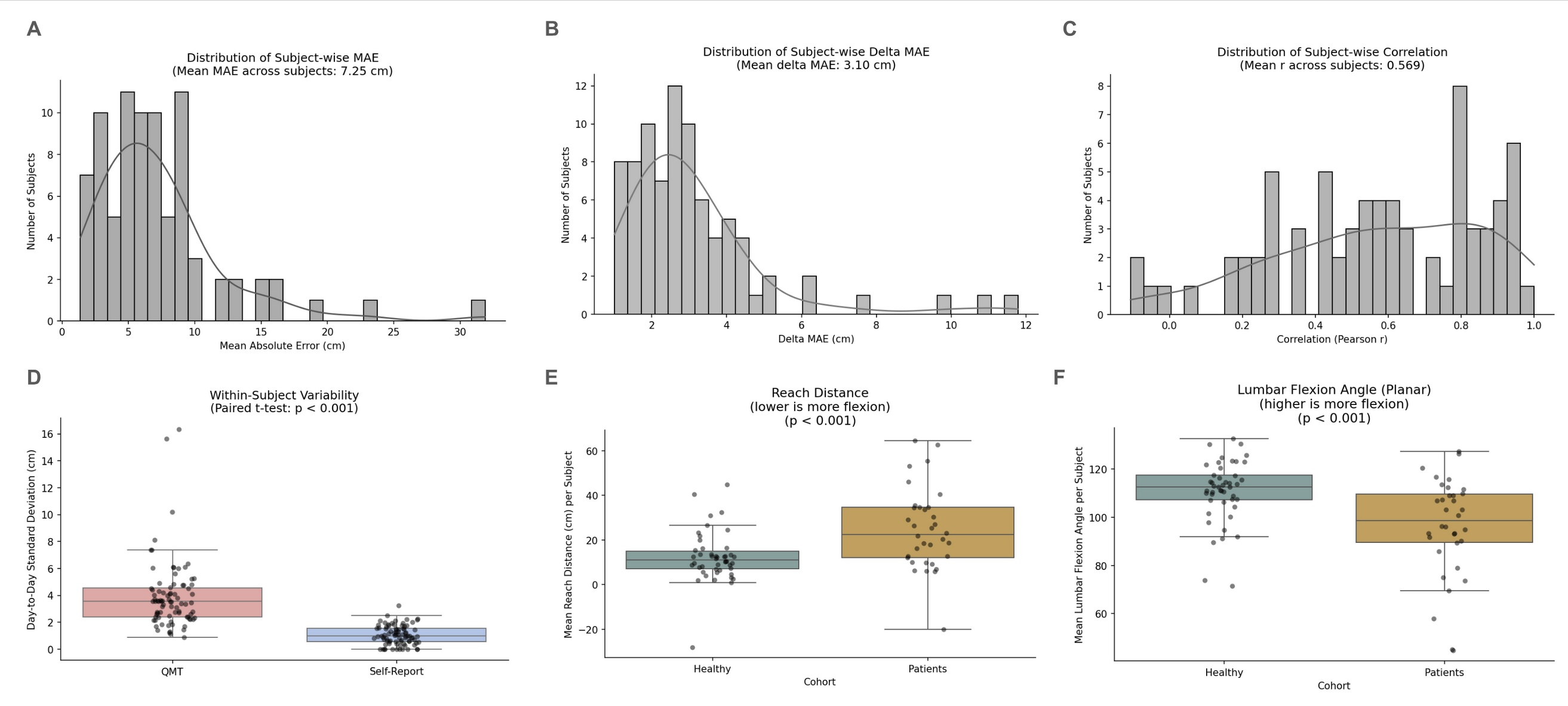} 
    \caption{\textbf{Agreement and Variability Between QMT and Subjective Self-Reports in the BeADS cohort.} (A) Distribution of subject-wise Mean Absolute Error (MAE) between calibrated QMT and self-reported reach distances. (B) Distribution of subject-wise $\Delta$ MAE between calibrated QMT and self-reported reach distances. $\Delta$ MAE measures how accurately QMT tracks the day-to-day change in a patient's symptoms, ignoring any constant baseline errors in the absolute measurements. (C) Distribution of subject-wise Pearson correlation ($r$) demonstrating moderate agreement between QMT and self-report trajectories. (D) Paired comparison of within-subject day-to-day variability (standard deviation). QMT measurements vary significantly more than subjective self-reports (paired t-test, $p < 0.001$). (E) Reach distance, measured as minimum wrist-to-ankle distance by QMT, is significantly lower in healthy participants compared to patients (independent t-test, $p < 0.001$). (F) Lumbar Flexion Angle (Planar) measured by QMT is significantly higher in healthy participants compared to patients (independent t-test, $p < 0.001$).}
    \label{fig:Fig4}
\end{figure}

\subsubsection*{Comparison to self-reports in a longitudinal study}

QMT reach estimates (wrist-to-ankle distance) were compared against concurrent patient self-reports (fingertip-to-floor distance) following the application of a uniform $-16.46$ cm calibration offset, derived using leave-one-subject-out cross-validation. 

The overall absolute agreement between the two modalities yielded a Mean Absolute Error (MAE) of 7.25 cm across all subjects (Figure \ref{fig:Fig4}A). Because home environments introduce variable recording setups that can cause fixed baseline offsets, we also evaluated the day-to-day change in measurements ($\Delta$ MAE). QMT tracked these day-to-day movement fluctuations with a mean $\Delta$ MAE of 3.10 cm (Figure \ref{fig:Fig4}B), indicating that while absolute distances may shift, the pipeline more reliably captures relative temporal changes, acknowledging that true day-to-day variance in this cohort was generally small. Furthermore, the longitudinal trajectories demonstrated a moderate intra-subject temporal relationship, with a mean Pearson correlation coefficient of $r = 0.569$ between QMT outputs and self-reported measures across the cohort (Figure \ref{fig:Fig4}C). Further, since more flexion means lower reach distance and higher lumbar flexion angle; therefore, they are inversely correlated (Supplementary Fig S6(B) and Fig. S7).  An analysis of day-to-day, within-subject variability revealed that QMT recorded significantly higher variance compared to the participants' corresponding self-reports (paired t-test, $p < 0.001$; Figure \ref{fig:Fig4}D). 

\subsubsection*{Cohort differentiation using QMT metrics}

Despite the increased noise of home recording, aggregating the longitudinal QMT metrics successfully differentiated the clinical cohorts. The minimum wrist-to-ankle reach distance measured by the QMT was significantly lower (indicating greater functional reach) in healthy controls compared to sciatica patients (independent t-test, $p < 0.001$; Figure \ref{fig:Fig4}E). Correspondingly, the planar lumbar flexion angle was significantly higher in the healthy cohort (independent t-test, $p < 0.001$; Figure \ref{fig:Fig4}F). A supplementary linear mixed-effects analysis confirmed the overall group difference (p<0.001). While a minor significant time effect was observed (p<0.001, capturing changes over the 30 days), the group × time interaction was not significant (Table S7). A few sample individual subject trajectories comparing calibrated QMT estimates and subjective self-reports over the 30-day monitoring period are visualised in Supplementary Figure S6.

Although these group-level differences help strengthen the construct validity of QMT, day-to-day metrics still exhibited significant overlap between healthy and patient groups. Therefore, when evaluating the discriminative capacity of these unaggregated daily metrics (i.e. reach distance and lumbar flexion angle on a single day) using a leave-one-subject-out logistic regression classifier, classification accuracy remained moderate to low (68.5\%, with area under ROC curve = 0.77, Figure S8). This discrepancy between highly significant group-level differences and limited single-day classification accuracy underscores the high intra-individual variance inherent to home monitoring, demonstrating that while the QMT captures valid digital biomarkers of impairment, robust individual stratification requires future work on utilising less noisy and a broad battery of measurements, along with sophisticated computational phenotyping.

\section*{Discussion}

This study outlines the development, protocol design, preliminary validation/benchmarking, and feasibility of prospective clinical deployment of Quantitative Movement Testing (QMT), a computer vision pipeline designed to balance the accessibility of monocular smartphone video with the clinical accuracy and interpretability of 3D kinematics. We establish a standardised chronic pain-relevant movement protocol and rigorously benchmark QMT against gold standard optical motion capture. Crucially, we then demonstrated the operational feasibility of this benchmarked pipeline across the translational spectrum: first, as a feasible secondary endpoint in a randomised controlled trial for fibromyalgia (PainLESS), and second, as a continuous at-home monitoring tool for patients with chronic sciatica (BeADS). QMT is a step towards making movement assessment equitable and accessible to diverse chronic pain populations, and this study evaluates its clinical utility, feasibility and reliability in the real world.

First, in the QMT validation study of 13 healthy subjects, we extracted clinical kinematic metrics with relatively high agreement to optical motion capture (correlations consistently exceeding $r=0.85$) after relevant bias. While we choose to present leave one subject out cross validated (LOSO-CV) results after calibration, we also provide uncalibrated results in Table 1 and in Supplementary figures. Calibration offsets derived from this validation study can be used in deployment (as a part of the QMT package) and future work needs to test its validity in out-of-distribution scenarios. We use VideoPose3D \citep{pavllo20193d} instead of more recent transformer-based 3D pose estimation models because it is one of the most extensively used and benchmarked models and is computationally light-weight, providing opportunities for deploying it on edge devices in the future \citep{lai2018rethinking}. A key contribution of our pipeline is the use of dynamic time warping (DTW) to segment and temporally align according to key predefined repeating movement motifs. DTW enables isolation of task-specific functional ranges. Furthermore, we found  was that planar (2D) surrogate angles significantly outperformed traditional 3D Euler derivations for trunk motion measurements, such as lumbar flexion. While QMT may not best capture 3D Euler angles in the way optimal motion capture does, QMT does a reasonable job of estimating motion in 2D planar surrogates. Notably, the mean absolute errors (MAE) observed for these planar angles (e.g., $2.7^\circ$ to $7.5^\circ$) perform comparably to standard clinical goniometry, which typically yields $\sim5^\circ$ of error \citep{macionis2013reliability} and significantly outperform clinical visual estimation, which frequently exhibits error margins of $9^\circ$--$12^\circ$ \citep{abu2010visual, cunha2020assessing}. Although the LOOCV-applied offsets were reasonably consistent within this sample (Supplementary Table~S1), the validation cohort was small, so these data should be interpreted as evidence that calibration improves group-level agreement rather than definitive proof that a fixed offset can be assumed for all future individuals across healthy and patient groups. Finally, the pipeline's diagnostic accuracy is reasonably maintained even when reducing the recording burden to three or a single trial.

Building on this technical foundation, the PainLESS trial demonstrated QMT's capability as a digital endpoint in a prospective clinical setting. QMT exhibited consistent intra-subject test-retest reliability across all evaluated metrics. The absence of detected differences between the dCBT-i and control group, is consistent with either a true null effect of dCBT-I on movement or with insufficient sensitivity of QMT to intervention-induced kinematic change. Without a positive control demonstrating responsiveness to known kinematic perturbations, these explanations cannot be disambiguated, and is an area for future work. The PainLESS data, therefore establish feasibility of kinematic capture in a clinical trial workflow, but responsiveness validation requires future work. 

Conversely, the longitudinal BeADS study highlighted both the construct validity and the practical limitations of translating this technology into unconstrained domestic environments. Despite the inherent challenges of remote recording, the pipeline successfully differentiated between healthy controls and patients with chronic sciatica based entirely on remote, self-captured video, establishing its value in identifying group-level differences capturing functional impairment. Importantly, the operational feasibility and patient acceptability of this remote tracking approach were encouraging; only two of the 45 enrolled patients withdrew from upload longitudinal videos, and 8.2\% of the unconstrained home recordings required manual exclusion due to severe framing errors. However, absolute measurement noise was approximately two to three times higher during at-home monitoring compared to the controlled validation study. This degradation is attributable to several factors. Environmentally, patients recorded movements under highly variable lighting conditions, at inconsistent distances and angles relative to their cameras, and occasionally with partial occlusions from furniture, other humans, or animals, and included reflections from mirrors. Methodologically, relying on a single daily trial intrinsically amplifies measurement variance compared to multi-trial laboratory averages. Lastly, higher variance in QMT measurements compared to self-reports may also partly stem from a psychological ``anchoring'' effect wherein participants artificially smooth their subjective reports by defaulting to previous days' scores.

These findings expose important limitations that direct our future work. First, our optical motion capture validation was conducted exclusively in a healthy controls (N=13) within a controlled laboratory setting. Future studies must validate the pipeline against optical motion capture by recreating noisy, home-like settings in the lab and testing variable recording angles and distances, directly within patient populations. Second, we deployed the VideoPose3D model in its pre-trained state; domain-specific fine-tuning of the underlying pose-estimation architecture on clinical movement datasets would likely yield further accuracy improvements. In summary, comparing the high fidelity observed in the PainLESS laboratory recordings to the variable data from the BeADS at-home study reveals that QMT currently performs most reliably within lab-based or clinic-based controlled environments. Nevertheless, developing intelligent user interfaces for real-time framing feedback and adjusting protocols to require three or more repetitions per home session could substantially smooth future algorithmic noise. 

Looking beyond remote monitoring, QMT can be deployed in a range of free-motion in-lab experiments involving experimentally induced pain and studying pain-related behaviour (e.g., \citet{tong2025phasic, mahajan2025balancing, eccleston2022prospective}). Ultimately, collecting high-frequency, objective kinematics opens promising avenues for predictive computational modelling \citep{mahajan2025forward, wiech2016deconstructing, jepma2018behavioural, mahajan2026homeostasis} to be applied outside of simple lab experiments to longitudinal real-world data. The utility of QMT in the PainLESS trial testing the effect of a sleep-based intervention highlights an avenue for studying the complex ways in which sleep and chronic pain affect each other \citep{kelleher2026sleep}. Furthermore, the feasibility of QMT and similar measurement tools provides a foundation for applying the aforementioned predictive computational modelling to potentially forecast which patients may recover and which may transition into worsening chronic pain states \citep{schmid2023factors}.

In conclusion, QMT demonstrates that monocular 3D pose estimation can serve as a scalable, objective tool for assessing functional movement relevant to chronic pain. While challenges remain in unconstrained domestic environments, this pipeline significantly lowers the operational barriers of traditional motion capture-based tracking, laying the groundwork for more accessible and equitable functional assessment in chronic pain populations.

\section*{Code Availability}

To support further benchmarking and community development, the complete QMT code pipeline is made openly available at: \href{https://github.com/PranavMahajan25/QMT}{https://github.com/PranavMahajan25/QMT}.

\section*{Acknowledgements}
The work was funded by Wellcome Trust (214251/Z/18/Z, 203139/Z/16/Z and 203139/A/16/Z), IITP (MSIT 2019-0-01371) and JSPS (22H04998). This research was also partly supported by the NIHR Oxford Health Biomedical Research Centre (NIHR203316), particularly the Pain Pump Priming (P3) Award. This grant is funded by UKRI and Versus Arthritis as part of the UKRI Strategic Priorities Fund (SPF) Advanced Pain Discovery Platform (APDP), a co-funded initiative by UKRI (MRC, BBSRC, ESRC), Versus Arthritis, the Medical Research Foundation and Eli Lilly and Company Ltd. The study was also supported by BMA Foundation Doris Hillier grant. A.I. is supported by the Margaret Harvey Schering Clinician Career Development Award Fund, the Daniel Foundation of Alabama Endowment at Mayo Clinic, Mayo Clinic Women’s Health Research Center, and Mayo Clinic Robert D. and Patricia E. Kern Center for the Science of Health Care Delivery. The views expressed are those of the author(s) and not necessarily those of the NIHR or the Department of Health and Social Care. For the purpose of open access, the authors have applied a CC BY public copyright licence to any Author Accepted Manuscript version arising from this submission.

\section*{Author Contributions}

\begin{table}[htbp]
\centering
\label{tab:author_contributions}
\begin{tabular}{ll}
\toprule
\textbf{Contributor Role} & \textbf{Author(s)} \\
\midrule
Conceptualisation & PM, AW, EC, BS \\
Data Curation & PM, AW, EC, EK \\
Formal Analysis & PM \\
Funding Acquisition & PM, AW, AS, KW, AI, BS \\
Investigation & PM, AW, EC  \\
Methodology & PM, AW, JS, AS, KW, AI \\
Project Administration & PM, AW, BS \\
Resources & JS \\
Software & PM, ST \\
Supervision & AS, KW, AI, BS \\
Validation & PM, AW \\
Visualization & PM, EK \\
Writing -- Original Draft Preparation & PM \\
Writing -- Review \& Editing & PM, AW, EC, JS, EK, ST, AS, KW, AI, BS \\
\bottomrule
\end{tabular}
\end{table}

\newpage

\bibliography{references}  

\newpage

\section*{Supplementary Materials}

\begin{figure}[htbp]
    \renewcommand\figurename{Figure S1}
    \renewcommand{\thefigure}{\hspace{-.333333em}}
    \centering
    \includegraphics[width=\textwidth]{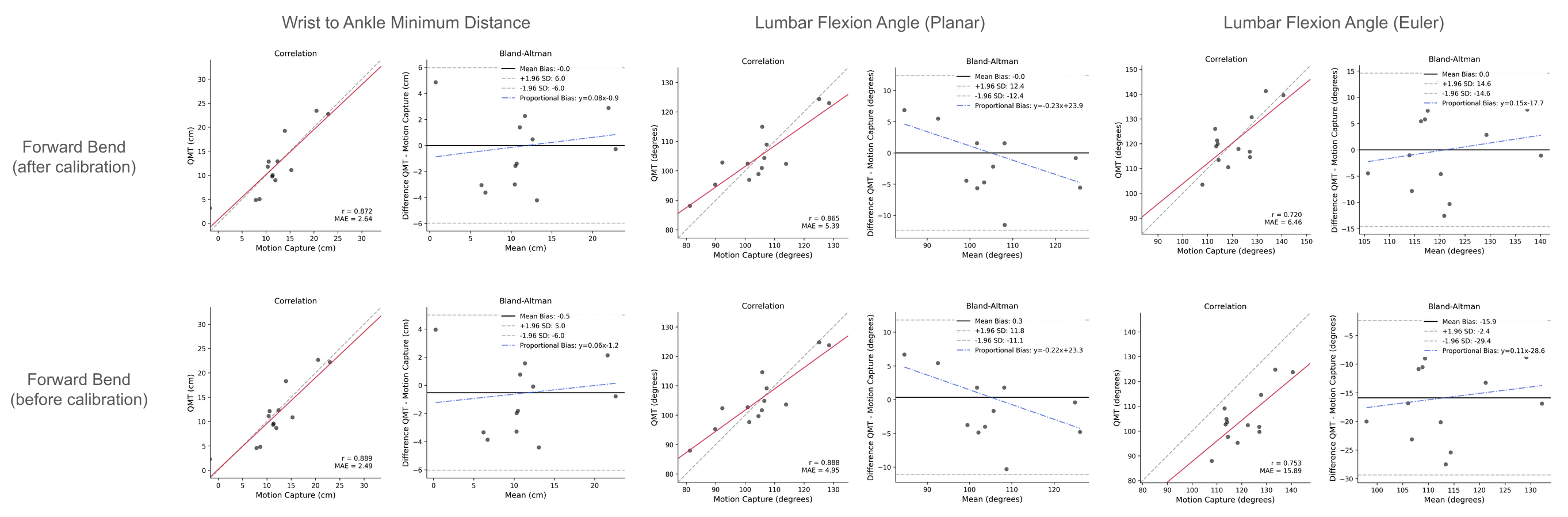} 
    \caption{Correlation and Bland-Altman plots of the Forward Bend movement (all metrics) before and after calibration. Using mean of 10 trials.}
    \label{fig:FigS1}
\end{figure}

\begin{figure}[htbp]
    \renewcommand\figurename{Figure S2}
    \renewcommand{\thefigure}{\hspace{-.333333em}}
    \centering
    \includegraphics[width=\textwidth]{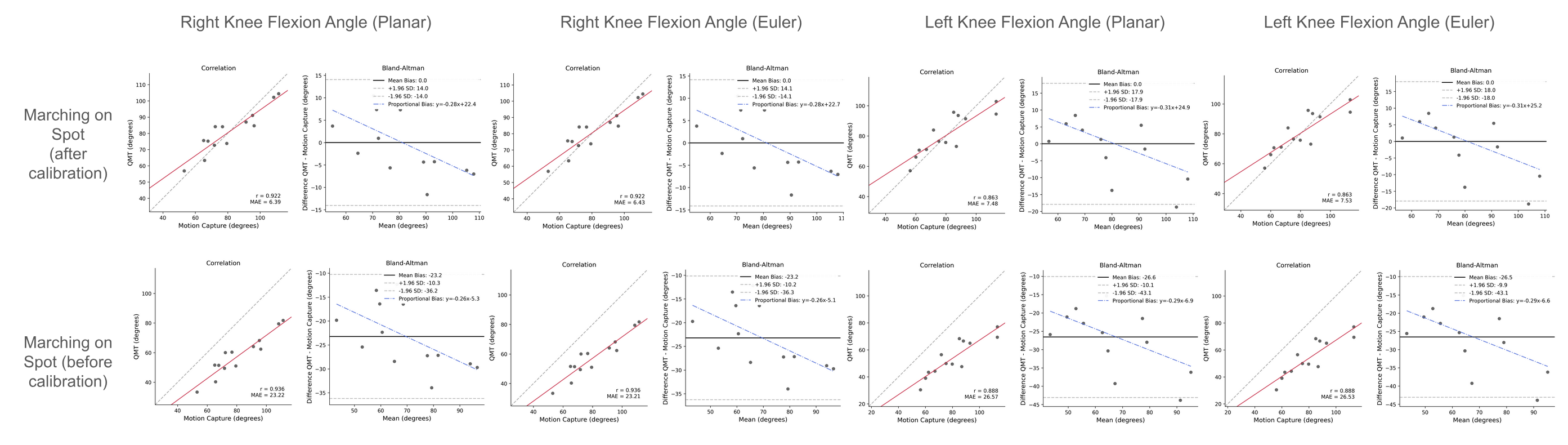} 
    \caption{Correlation and Bland-Altman plots of the Marching on Spot movement (knee metrics) before and after calibration. Using mean of 10 trials.}
    \label{fig:FigS2}
\end{figure}

\begin{figure}[htbp]
    \renewcommand\figurename{Figure S3}
    \renewcommand{\thefigure}{\hspace{-.333333em}}
    \centering
    \includegraphics[width=\textwidth]{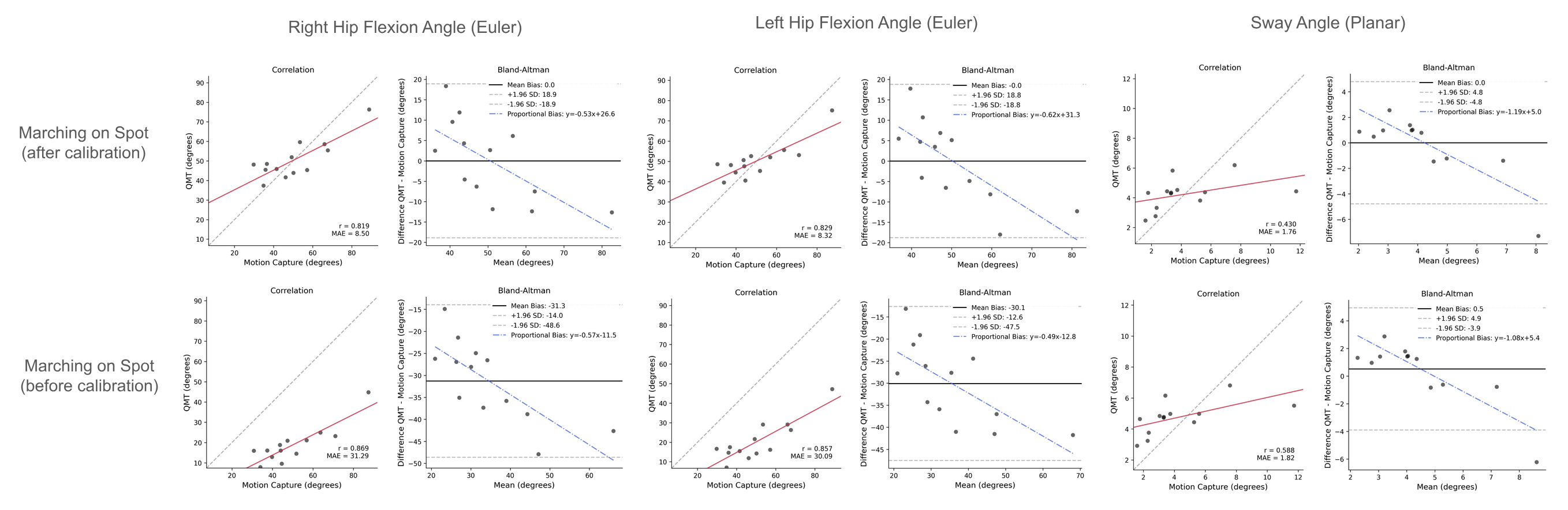} 
    \caption{Correlation and Bland-Altman plots of the Marching on Spot movement (hip and sway metrics) before and after calibration. Using mean of 10 trials.}
    \label{fig:FigS3}
\end{figure}

\begin{figure}[htbp]
    \renewcommand\figurename{Figure S4}
    \renewcommand{\thefigure}{\hspace{-.333333em}}
    \centering
    \includegraphics[width=\textwidth]{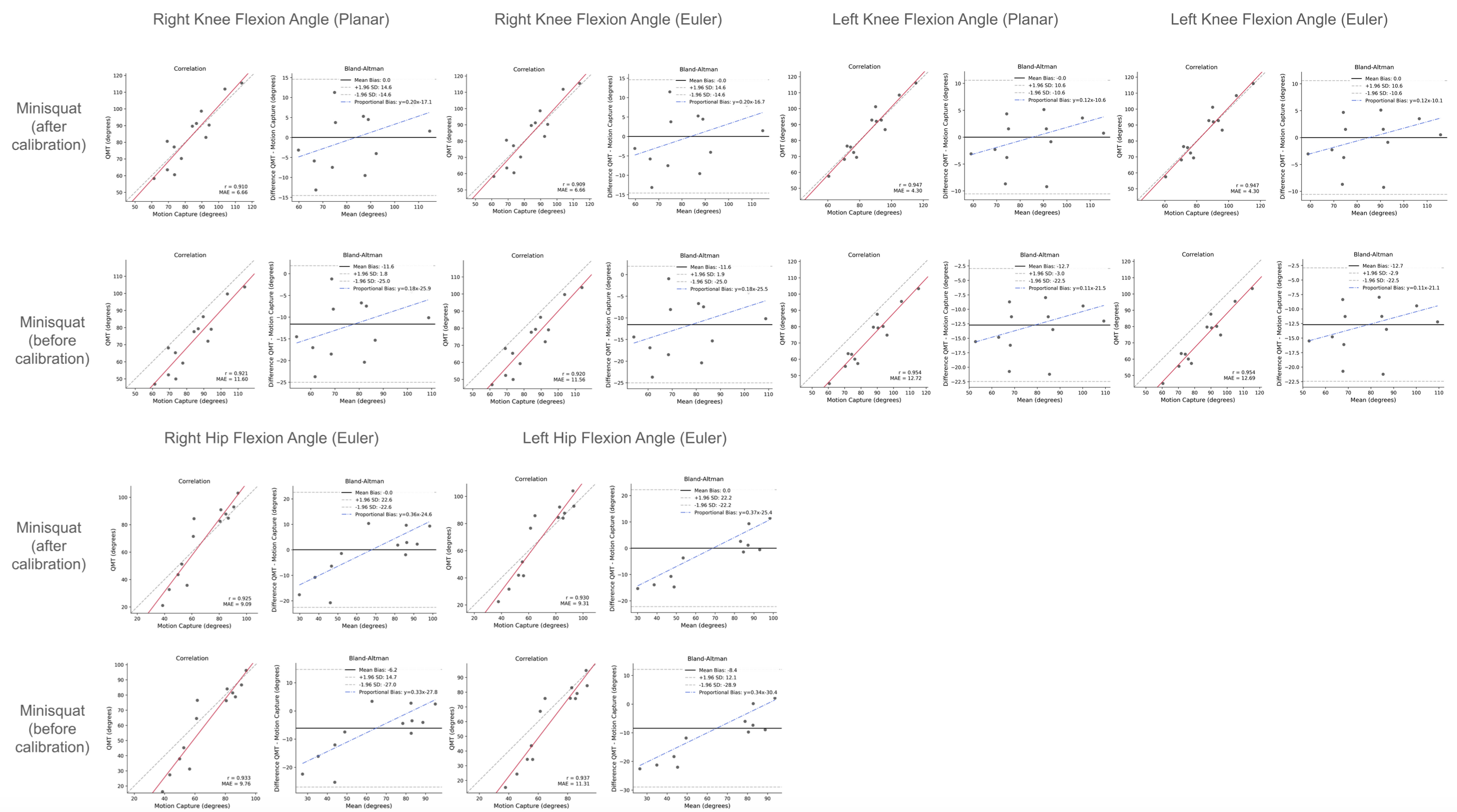} 
    \caption{Correlation and Bland-Altman plots of the Minisquat movement (knee and hip metrics) before and after calibration. Using mean of 10 trials.}
    \label{fig:FigS4}
\end{figure}

\begin{figure}[htbp]
    \renewcommand\figurename{Figure S5}
    \renewcommand{\thefigure}{\hspace{-.333333em}}
    \centering
    \includegraphics[width=\textwidth]{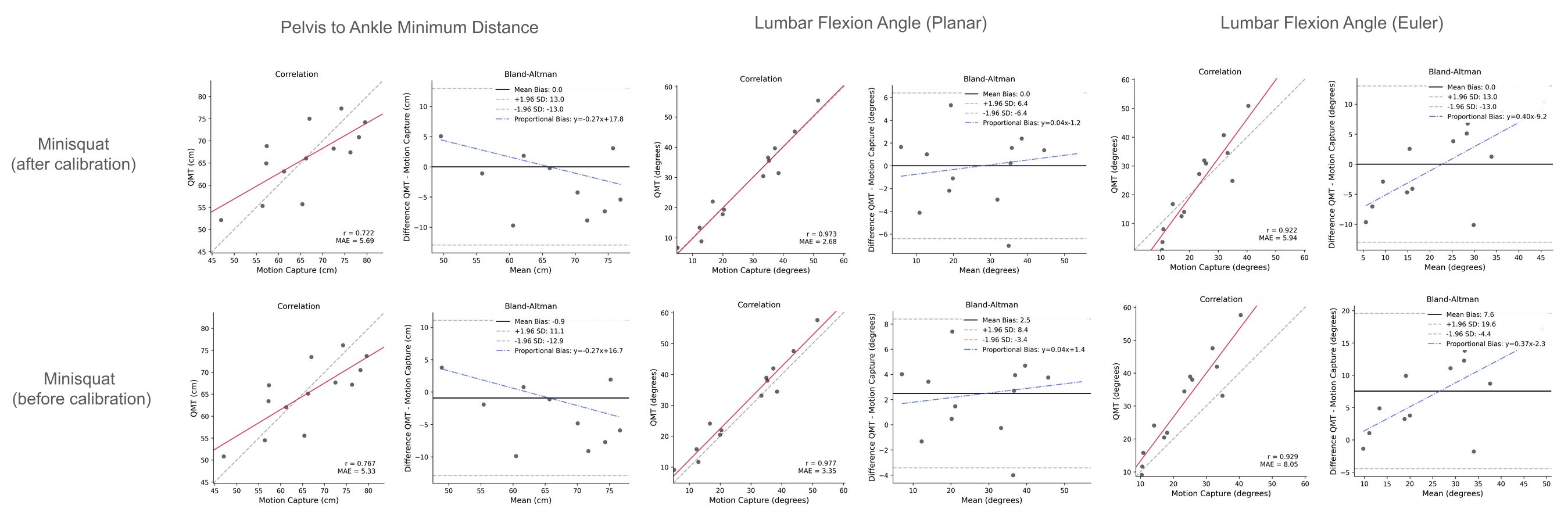} 
    \caption{Correlation and Bland-Altman plots of the Minisquat movement (remaining metrics) before and after calibration. Using mean of 10 trials.}
    \label{fig:FigS5}
\end{figure}

\newpage
\begin{table}[htbp]
\renewcommand\tablename{Table S1}
\renewcommand{\thetable}{\hspace{-.333333em}}
\centering
\caption{Optical motion-capture markers used in the laboratory validation protocol, including anatomical landmarks and placement descriptions. Marker placement followed the Oxford Gait Laboratory Plug-in-Gait protocol with additional task-specific markers used locally where noted.}
\label{tab:marker_descriptions}
\resizebox{\textwidth}{!}{
\begin{tabular}{llll}
\toprule
\textbf{Marker Group} & \textbf{Marker} & \textbf{Site} & \textbf{Marker Placement} \\
\midrule
Torso & T2 & 2nd thoracic vertebra & Spinous process of the 2nd thoracic vertebra \\
Torso & T6 & 6th thoracic vertebra & Spinous process of the 6th thoracic vertebra \\
Torso & T12 & 12th thoracic vertebra & Spinous process of the 12th thoracic vertebra \\
\midrule
Pelvis & LPSI & Left PSIS & Placed directly over the left posterior superior iliac spine \\
Pelvis & RPSI & Right PSIS & Placed directly over the right posterior superior iliac spine \\
Pelvis & SACR & Sacrum & Placed midway between the posterior superior iliac spines \\
Pelvis & LASI & Left ASIS & Placed directly over the left anterior superior iliac spine \\
Pelvis & RASI & Right ASIS & Placed directly over the right anterior superior iliac spine \\
\midrule
Leg & LTHI & Left thigh & Placed over the lower lateral one-third surface of the left thigh, just below hand swing \\
Leg & RTHI & Right thigh & Placed over the lower lateral one-third surface of the right thigh, just below hand swing \\
Leg & LTHA1 & Left upper thigh & Placed at the first one-third point of the left anterior thigh \\
Leg & LTHA2 & Left lower thigh & Placed at the second one-third point of the left anterior thigh \\
Leg & LTHL & Left lateral thigh & Placed on the lower one-third surface of the left lateral thigh \\
Leg & RTHA1 & Right upper thigh & Placed at the first one-third point of the right anterior thigh \\
Leg & RTHA2 & Right lower thigh & Placed at the second one-third point of the right anterior thigh \\
Leg & RTHL & Right lateral thigh & Placed on the lower one-third surface of the right lateral thigh \\
Leg & LKNE & Left knee & Placed on the lateral epicondyle of the left knee \\
Leg & RKNE & Right knee & Placed on the lateral epicondyle of the right knee \\
Leg & LHFB & Left head of fibula & Placed on the most lateral aspect of the left fibular head \\
Leg & RHFB & Right head of fibula & Placed on the most lateral aspect of the right fibular head \\
Leg & LTUB & Left tibial tuberosity & Placed on the most anterior aspect of the left tibial tuberosity \\
Leg & RTUB & Right tibial tuberosity & Placed on the most anterior aspect of the right tibial tuberosity \\
Leg & LSHN & Left shin & Placed on the anterior crest of the left tibia \\
Leg & RSHN & Right shin & Placed on the anterior crest of the right tibia \\
Leg & LANK & Left lateral ankle & Placed on the lateral malleolus along the transmalleolar axis \\
Leg & RANK & Right lateral ankle & Placed on the lateral malleolus along the transmalleolar axis \\
Leg & LMMA & Left medial ankle & Placed on the medial malleolus along the transmalleolar axis \\
Leg & RMMA & Right medial ankle & Placed on the medial malleolus along the transmalleolar axis \\
\midrule
Arm & LFIN & Left finger & Placed on the nailbed of the left middle finger \\
Arm & RFIN & Right finger & Placed on the nailbed of the right middle finger \\
Arm & LWR & Left wrist & Placed in the middle of the back of the left wrist, on the joint line \\
Arm & RWR & Right wrist & Placed in the middle of the back of the right wrist, on the joint line \\
\bottomrule
\end{tabular}
}
\end{table}

\begin{table}[htbp]
\renewcommand\tablename{Table S2}
\renewcommand{\thetable}{\hspace{-.333333em}}
\centering
\caption{Participant-specific leave-one-subject-out calibration offsets applied for the supplementary movement-metric subset. Global offset is the mean of the subject-specific applied offsets.}
\label{tab:calibration_subject_offsets}
\resizebox{\textwidth}{!}{%
\begin{tabular}{llrrrrrrrrrrrrrrrrrr}
\toprule
\textbf{Movement} & \textbf{Metric} & \textbf{N} & \textbf{Raw Bias} & \textbf{QMT01} & \textbf{QMT02} & \textbf{QMT03} & \textbf{QMT04} & \textbf{QMT05} & \textbf{QMT06} & \textbf{QMT07} & \textbf{QMT08} & \textbf{QMT09} & \textbf{QMT10} & \textbf{QMT11} & \textbf{QMT12} & \textbf{QMT13} & \textbf{Global Offset} & \textbf{Offset SD} & \textbf{Offset Range} \\
\midrule
Forward Bend & Wrist to Ankle Min Dist & 13 & -0.53 & 0.56 & 0.50 & 0.93 & 0.29 & 0.42 & 0.90 & 0.20 & 0.29 & 0.70 & 0.40 & 0.63 & 0.75 & 0.25 & 0.53 & 0.24 & 0.73 \\
Forward Bend & Lumbar Flexion (Euler) & 13 & -15.89 & 14.92 & 15.28 & 15.81 & 16.10 & 16.88 & 15.80 & 15.53 & 16.46 & 15.09 & 16.30 & 16.33 & 15.54 & 16.48 & 15.89 & 0.60 & 1.96 \\
Forward Bend & Lumbar Flexion (Planar) & 13 & 0.35 & -0.78 & 0.07 & -0.69 & 0.35 & -0.23 & -0.78 & 0.46 & -0.52 & -0.71 & -0.23 & -1.24 & 0.18 & -0.41 & -0.35 & 0.51 & 1.70 \\
\midrule
Marching on Spot & Right Knee Flexion (Euler) & 13 & -23.21 & 22.31 & 22.87 & 22.72 & 23.28 & 23.77 & 24.12 & 23.50 & 22.66 & 24.01 & 22.88 & 22.78 & 23.03 & 23.77 & 23.21 & 0.58 & 1.82 \\
Marching on Spot & Right Knee Flexion (Planar) & 13 & -23.22 & 22.33 & 22.89 & 22.74 & 23.29 & 23.78 & 24.13 & 23.50 & 22.68 & 24.02 & 22.89 & 22.79 & 23.04 & 23.79 & 23.22 & 0.57 & 1.80 \\
Marching on Spot & Left Knee Flexion (Euler) & 13 & -26.53 & 26.40 & 27.34 & 25.08 & 26.63 & 25.47 & 27.48 & 26.61 & 25.73 & 26.84 & 26.95 & 26.21 & 26.99 & 27.18 & 26.53 & 0.73 & 2.39 \\
Marching on Spot & Left Knee Flexion (Planar) & 13 & -26.57 & 26.45 & 27.38 & 25.13 & 26.67 & 25.51 & 27.50 & 26.62 & 25.77 & 26.88 & 26.99 & 26.25 & 27.02 & 27.21 & 26.57 & 0.73 & 2.38 \\
Marching on Spot & Right Hip Flexion (Euler) & 13 & -31.29 & 31.69 & 31.82 & 29.91 & 30.79 & 31.56 & 32.12 & 31.72 & 30.35 & 30.98 & 30.92 & 30.67 & 32.66 & 31.66 & 31.29 & 0.77 & 2.75 \\
Marching on Spot & Left Hip Flexion (Euler) & 13 & -30.09 & 30.29 & 29.74 & 29.51 & 29.18 & 29.60 & 31.00 & 30.28 & 29.12 & 30.42 & 30.56 & 29.14 & 31.50 & 30.82 & 30.09 & 0.77 & 2.38 \\
Marching on Spot & Side Sway (Planar) & 13 & 0.51 & -0.61 & -0.63 & -0.62 & -0.44 & -0.44 & -0.41 & -0.45 & -1.08 & -0.32 & -0.33 & -0.45 & -0.48 & -0.44 & -0.51 & 0.20 & 0.76 \\
\midrule
Minisquat & Pelvis to Ankle Min Dist & 13 & -0.91 & 1.05 & 0.50 & 1.53 & 1.30 & 0.90 & 1.50 & 0.23 & 0.83 & 1.80 & 0.17 & 0.59 & 0.35 & 1.15 & 0.91 & 0.53 & 1.63 \\
Minisquat & Lumbar Flexion (Euler) & 13 & 7.56 & -8.34 & -8.30 & -6.89 & -7.27 & -7.46 & -7.17 & -7.36 & -7.04 & -6.76 & -7.88 & -8.10 & -7.92 & -7.78 & -7.56 & 0.53 & 1.58 \\
Minisquat & Lumbar Flexion (Planar) & 13 & 2.48 & -2.71 & -2.36 & -2.38 & -3.02 & -2.30 & -2.47 & -2.07 & -2.36 & -2.18 & -2.57 & -2.80 & -2.65 & -2.41 & -2.48 & 0.26 & 0.95 \\
Minisquat & Right Knee Flexion (Euler) & 13 & -11.56 & 11.90 & 10.55 & 11.85 & 11.67 & 11.97 & 10.82 & 12.44 & 12.18 & 11.25 & 12.24 & 11.11 & 10.98 & 11.32 & 11.56 & 0.60 & 1.89 \\
Minisquat & Right Knee Flexion (Planar) & 13 & -11.60 & 11.95 & 10.59 & 11.89 & 11.72 & 12.01 & 10.87 & 12.47 & 12.22 & 11.29 & 12.29 & 11.15 & 11.02 & 11.36 & 11.60 & 0.59 & 1.88 \\
Minisquat & Left Knee Flexion (Euler) & 13 & -12.69 & 12.81 & 12.81 & 12.40 & 12.73 & 13.08 & 11.98 & 13.05 & 12.96 & 12.62 & 13.54 & 12.51 & 12.02 & 12.45 & 12.69 & 0.43 & 1.57 \\
Minisquat & Left Knee Flexion (Planar) & 13 & -12.72 & 12.84 & 12.84 & 12.43 & 12.77 & 13.11 & 12.01 & 13.05 & 12.99 & 12.65 & 13.57 & 12.54 & 12.05 & 12.48 & 12.72 & 0.43 & 1.57 \\
Minisquat & Right Hip Flexion (Euler) & 13 & -6.15 & 7.90 & 4.79 & 6.00 & 6.89 & 6.37 & 6.29 & 6.04 & 6.32 & 6.87 & 6.95 & 4.55 & 5.66 & 5.32 & 6.15 & 0.92 & 3.35 \\
Minisquat & Left Hip Flexion (Euler) & 13 & -8.42 & 10.04 & 7.24 & 8.31 & 9.14 & 8.51 & 8.62 & 8.14 & 8.38 & 9.30 & 9.58 & 7.29 & 7.60 & 7.35 & 8.42 & 0.91 & 2.80 \\
\bottomrule
\end{tabular}%
}
\end{table}

\begin{table}[htbp]
\renewcommand\tablename{Table S3}
\renewcommand{\thetable}{\hspace{-.333333em}}
\centering
\caption{Comparison of Results (After Calibration) Using Mean of 10, 3, and 1 Trials}
\label{tab:reps_comparison}
\resizebox{\textwidth}{!}{
\begin{tabular}{llrrrrrrrrrrrrrrr}
\toprule
& & \multicolumn{5}{c}{\textbf{10 Trials}} & \multicolumn{5}{c}{\textbf{3 Trials}} & \multicolumn{5}{c}{\textbf{1 Trial}} \\
\cmidrule(lr){3-7} \cmidrule(lr){8-12} \cmidrule(lr){13-17}
\textbf{Movement} & \textbf{Metric} & \textbf{Bias} & \textbf{LoA Lower} & \textbf{LoA Upper} & \textbf{r} & \textbf{MAE} & \textbf{Bias} & \textbf{LoA Lower} & \textbf{LoA Upper} & \textbf{r} & \textbf{MAE} & \textbf{Bias} & \textbf{LoA Lower} & \textbf{LoA Upper} & \textbf{r} & \textbf{MAE} \\
\midrule
Forward Bend & Wrist to Ankle Min Dist & 0.00 & -5.98 & 5.98 & 0.87 & 2.64 & 1.25 & -6.34 & 8.83 & 0.88 & 3.55 & 2.48 & -5.11 & 10.07 & 0.89 & 3.93 \\
 & Lumbar Flexion (Euler) & 0.00 & -14.60 & 14.60 & 0.72 & 6.46 & -1.06 & -18.26 & 16.14 & 0.70 & 7.45 & -2.02 & -18.94 & 14.90 & 0.72 & 6.92 \\
 & Lumbar Flexion (Planar) & 0.00 & -12.40 & 12.40 & 0.86 & 5.39 & -1.06 & -12.89 & 10.77 & 0.88 & 5.24 & -2.02 & -14.20 & 10.17 & 0.87 & 5.70 \\
\midrule
Marching on Spot & Right Knee Flexion (Euler) & 0.00 & -14.14 & 14.14 & 0.92 & 6.43 & -3.61 & -19.64 & 12.42 & 0.88 & 8.14 & -5.36 & -20.77 & 10.06 & 0.89 & 8.78 \\
 & Right Knee Flexion (Planar) & 0.00 & -14.05 & 14.05 & 0.92 & 6.39 & -3.61 & -19.56 & 12.34 & 0.88 & 8.10 & -5.36 & -20.70 & 9.99 & 0.89 & 8.75 \\
 & Left Knee Flexion (Euler) & 0.00 & -17.97 & 17.97 & 0.86 & 7.53 & -3.90 & -21.15 & 13.34 & 0.90 & 6.69 & -8.85 & -27.96 & 10.25 & 0.85 & 10.61 \\
 & Left Knee Flexion (Planar) & 0.00 & -17.89 & 17.89 & 0.86 & 7.48 & -3.90 & -21.05 & 13.25 & 0.90 & 6.63 & -8.85 & -27.86 & 10.15 & 0.85 & 10.56 \\
 & Right Hip Flexion (Euler) & 0.00 & -18.78 & 18.78 & 0.83 & 8.32 & -0.89 & -23.07 & 21.29 & 0.72 & 9.36 & -1.43 & -28.38 & 25.51 & 0.45 & 11.14 \\
 & Left Hip Flexion (Euler) & 0.00 & -18.90 & 18.90 & 0.82 & 8.50 & -0.86 & -20.92 & 19.21 & 0.82 & 8.56 & -1.83 & -24.24 & 20.59 & 0.78 & 8.75 \\
 & Side Sway (Planar) & 0.00 & -4.79 & 4.79 & 0.43 & 1.76 & -0.04 & -5.38 & 5.29 & 0.22 & 1.87 & -0.33 & -6.43 & 5.77 & 0.10 & 2.22 \\
\midrule
Minisquat & Pelvis to Ankle Min Dist & 0.00 & -12.96 & 12.96 & 0.72 & 5.69 & 1.24 & -12.34 & 14.82 & 0.69 & 6.42 & 2.85 & -11.14 & 16.85 & 0.66 & 7.01 \\
 & Lumbar Flexion (Euler) & 0.00 & -13.01 & 13.01 & 0.92 & 5.94 & -0.79 & -13.73 & 12.14 & 0.89 & 5.97 & -1.29 & -13.97 & 11.38 & 0.87 & 5.49 \\
 & Lumbar Flexion (Planar) & 0.00 & -6.40 & 6.40 & 0.97 & 2.68 & -0.79 & -9.56 & 7.97 & 0.94 & 3.53 & -1.29 & -11.67 & 9.08 & 0.92 & 4.31 \\
 & Right Knee Flexion (Euler) & 0.00 & -14.59 & 14.59 & 0.91 & 6.66 & -1.74 & -16.02 & 12.55 & 0.89 & 5.85 & -4.44 & -19.83 & 10.95 & 0.86 & 7.21 \\
 & Right Knee Flexion (Planar) & 0.00 & -14.56 & 14.56 & 0.91 & 6.66 & -1.74 & -15.96 & 12.48 & 0.89 & 5.85 & -4.44 & -19.74 & 10.87 & 0.86 & 7.18 \\
 & Left Knee Flexion (Euler) & 0.00 & -10.59 & 10.59 & 0.95 & 4.30 & -1.39 & -14.17 & 11.40 & 0.91 & 5.27 & -4.03 & -17.74 & 9.68 & 0.89 & 6.07 \\
 & Left Knee Flexion (Planar) & 0.00 & -10.56 & 10.56 & 0.95 & 4.30 & -1.39 & -14.15 & 11.38 & 0.91 & 5.26 & -4.03 & -17.71 & 9.65 & 0.89 & 6.06 \\
 & Right Hip Flexion (Euler) & 0.00 & -22.61 & 22.61 & 0.93 & 9.09 & -2.04 & -25.35 & 21.28 & 0.89 & 9.90 & -4.06 & -27.34 & 19.21 & 0.87 & 9.96 \\
 & Left Hip Flexion (Euler) & 0.00 & -22.23 & 22.23 & 0.93 & 9.31 & -2.08 & -25.83 & 21.68 & 0.89 & 10.44 & -4.34 & -29.11 & 20.44 & 0.87 & 10.85 \\
\bottomrule
\end{tabular}
}
\end{table}

\begin{table}[htbp]
\renewcommand\tablename{Table S4}
\renewcommand{\thetable}{\hspace{-.333333em}}
\centering
\caption{Within-Group Secondary Analysis of Pre- and Post-Intervention Changes (PainLESS Trial)}
\label{tab:within_painless}
\resizebox{\textwidth}{!}{
\begin{tabular}{lllrrrrrrrr}
\toprule
\textbf{Movement} & \textbf{Metric} & \textbf{Group} & \textbf{N} & \textbf{Pre Mean} & \textbf{Post Mean} & \textbf{$\Delta$ Mean} & \textbf{Paired t} & \textbf{p-value} & \textbf{BF$_{10}$} & \textbf{Evidence} \\
\midrule
Forward Bend & Wrist to Ankle Min Dist (cm) & dCBT-I & 28 & 27.63 & 24.86 & -2.77 & -1.01 & 0.321 & 0.32 & Moderate $H_0$ \\
  &   & Control & 26 & 21.45 & 16.98 & -4.48 & -1.11 & 0.276 & 0.36 & Anecdotal $H_0$ \\
  & Lumbar Flexion (Planar) & dCBT-I & 28 & 89.40 & 93.06 & 3.67 & 1.32 & 0.199 & 0.44 & Anecdotal $H_0$ \\
  &   & Control & 26 & 97.51 & 101.11 & 3.60 & 1.02 & 0.319 & 0.33 & Moderate $H_0$ \\
\midrule
Marching on Spot & Right Knee Flexion (Planar) & dCBT-I & 30 & 70.58 & 72.69 & 2.11 & 0.79 & 0.434 & 0.26 & Moderate $H_0$ \\
  &   & Control & 27 & 73.63 & 73.19 & -0.44 & -0.12 & 0.904 & 0.21 & Moderate $H_0$ \\
  & Left Knee Flexion (Planar) & dCBT-I & 30 & 72.32 & 71.49 & -0.83 & -0.28 & 0.780 & 0.20 & Moderate $H_0$ \\
  &   & Control & 27 & 76.78 & 71.82 & -4.95 & -1.41 & 0.171 & 0.49 & Anecdotal $H_0$ \\
  & Right Hip Flexion (Planar) & dCBT-I & 30 & 37.42 & 44.42 & 7.00 & 1.64 & 0.112 & 0.64 & Anecdotal $H_0$ \\
  &   & Control & 27 & 39.57 & 39.02 & -0.55 & -0.15 & 0.878 & 0.21 & Moderate $H_0$ \\
  & Left Hip Flexion (Planar) & dCBT-I & 30 & 34.55 & 38.91 & 4.35 & 1.12 & 0.270 & 0.35 & Anecdotal $H_0$ \\
  &   & Control & 27 & 38.63 & 35.02 & -3.61 & -1.10 & 0.281 & 0.35 & Anecdotal $H_0$ \\
  & Side Sway (Planar) & dCBT-I & 30 & 6.13 & 10.51 & 4.37 & 1.88 & 0.070 & 0.92 & Anecdotal $H_0$ \\
  &   & Control & 27 & 6.63 & 6.71 & 0.08 & 0.15 & 0.879 & 0.21 & Moderate $H_0$ \\
\midrule
Minisquat & Pelvis to Ankle Min Dist (cm) & dCBT-I & 30 & 63.56 & 59.61 & -3.95 & -1.00 & 0.325 & 0.31 & Moderate $H_0$ \\
  &   & Control & 27 & 60.82 & 59.34 & -1.47 & -0.48 & 0.632 & 0.23 & Moderate $H_0$ \\
  & Lumbar Flexion (Planar) & dCBT-I & 30 & 27.90 & 37.08 & 9.18 & 1.72 & 0.096 & 0.72 & Anecdotal $H_0$ \\
  &   & Control & 27 & 30.96 & 33.90 & 2.94 & 1.11 & 0.275 & 0.36 & Anecdotal $H_0$ \\
  & Right Knee Flexion (Planar) & dCBT-I & 30 & 65.01 & 68.50 & 3.49 & 0.77 & 0.447 & 0.26 & Moderate $H_0$ \\
  &   & Control & 27 & 68.98 & 72.22 & 3.24 & 0.70 & 0.491 & 0.25 & Moderate $H_0$ \\
  & Left Knee Flexion (Planar) & dCBT-I & 30 & 67.18 & 69.58 & 2.40 & 0.62 & 0.543 & 0.23 & Moderate $H_0$ \\
  &   & Control & 27 & 72.21 & 72.63 & 0.42 & 0.11 & 0.915 & 0.20 & Moderate $H_0$ \\
  & Right Hip Flexion (Planar) & dCBT-I & 30 & 45.33 & 54.16 & 8.84 & 1.95 & 0.061 & 1.02 & Anecdotal $H_1$ \\
  &   & Control & 27 & 57.34 & 56.35 & -0.99 & -0.20 & 0.840 & 0.21 & Moderate $H_0$ \\
  & Left Hip Flexion (Planar) & dCBT-I & 30 & 43.38 & 53.35 & 9.97 & 2.21 & 0.035 & 1.59 & Anecdotal $H_1$ \\
  &   & Control & 27 & 56.08 & 55.10 & -0.99 & -0.21 & 0.833 & 0.21 & Moderate $H_0$ \\
\bottomrule
\end{tabular}
}
\end{table}

\begin{table}[htbp]
\renewcommand\tablename{Table S5}
\renewcommand{\thetable}{\hspace{-.333333em}}
\centering
\caption{Mixed-Effects Model Analysis for Group $\times$ Time Interaction (PainLESS Trial)}
\label{tab:mixed_model_painless}
\resizebox{\textwidth}{!}{
\begin{tabular}{llrrrrrr}
\toprule
\textbf{Movement} & \textbf{Metric} & \textbf{N} & \textbf{Interaction LR} & \textbf{df} & \textbf{p-value} & \textbf{BF$_{10}$} & \textbf{Evidence} \\
\midrule
Forward Bend & Wrist to Ankle Min Dist (cm) & 54 & 0.13 & 1 & 0.717 & 0.10 & Moderate $H_0$ \\
  & Lumbar Flexion (Planar) & 54 & 0.00 & 1 & 0.988 & 0.10 & Strong $H_0$ \\
\midrule
Marching on Spot & Right Knee Flexion (Planar) & 57 & 0.34 & 1 & 0.557 & 0.11 & Moderate $H_0$ \\
  & Left Knee Flexion (Planar) & 57 & 0.85 & 1 & 0.358 & 0.14 & Moderate $H_0$ \\
  & Right Hip Flexion (Planar) & 57 & 1.84 & 1 & 0.175 & 0.24 & Moderate $H_0$ \\
  & Left Hip Flexion (Planar) & 57 & 2.44 & 1 & 0.118 & 0.32 & Moderate $H_0$ \\
  & Side Sway (Planar) & 57 & 2.99 & 1 & 0.084 & 0.42 & Anecdotal $H_0$ \\
\midrule
Minisquat & Pelvis to Ankle Min Dist (cm) & 57 & 0.25 & 1 & 0.619 & 0.11 & Moderate $H_0$ \\
  & Lumbar Flexion (Planar) & 57 & 1.05 & 1 & 0.305 & 0.16 & Moderate $H_0$ \\
  & Right Knee Flexion (Planar) & 57 & 0.00 & 1 & 0.969 & 0.09 & Strong $H_0$ \\
  & Left Knee Flexion (Planar) & 57 & 0.13 & 1 & 0.718 & 0.10 & Strong $H_0$ \\
  & Right Hip Flexion (Planar) & 57 & 2.22 & 1 & 0.136 & 0.28 & Moderate $H_0$ \\
  & Left Hip Flexion (Planar) & 57 & 2.90 & 1 & 0.089 & 0.40 & Anecdotal $H_0$ \\
\bottomrule
\end{tabular}
}
\end{table}

\begin{table}[htbp]
\renewcommand\tablename{Table S6}
\renewcommand{\thetable}{\hspace{-.333333em}}
\centering
\caption{Primary Group Comparison of Pre- to Post-Intervention Changes (PainLESS Trial - Compliant Sleepio vs Control). 
Participants who completed 3 or more sessions of dCBT-I were termed as compliant (while those who did fewer were filtered out). N (dCBT-I / Ctrl) refers to the number of participants in the intervention group who were compliant and the control group, respectively. $\Delta$ denotes the mean change from baseline to post-intervention (post-intervention minus pre-intervention). Evidence strength is based on the Bayes Factor ($BF_{10}$).}
\label{tab:primary_painless_compliant}
\resizebox{\textwidth}{!}{
\begin{tabular}{llrrrrrrrrr}
\toprule
\textbf{Movement} & \textbf{Metric} & \textbf{N (dCBT-I / Ctrl)} & \textbf{$\Delta$ dCBT-I} & \textbf{$\Delta$ Ctrl} & \textbf{Diff ($\Delta$)} & \textbf{Welch's t} & \textbf{p-value} & \textbf{Hedges' $g$} & \textbf{BF$_{10}$} & \textbf{Evidence} \\
\midrule
Forward Bend & Wrist to Ankle Min Dist (cm) & 16 / 26 & -4.25 & -4.48 & 0.23 & 0.04 & 0.969 & 0.01 & 0.15 & Moderate $H_0$ \\
~ & Lumbar Flexion (Planar) & 16 / 26 & 6.23 & 3.60 & 2.63 & 0.46 & 0.649 & 0.14 & 0.17 & Moderate $H_0$ \\
\midrule
Marching on Spot & Left Knee Flexion (Planar) & 17 / 27 & 0.46 & -4.95 & 5.41 & 1.14 & 0.262 & 0.32 & 0.27 & Moderate $H_0$ \\
~ & Right Knee Flexion (Planar) & 17 / 27 & 2.01 & -0.44 & 2.45 & 0.54 & 0.595 & 0.15 & 0.17 & Moderate $H_0$ \\
~ & Left Hip Flexion (Planar) & 17 / 27 & 8.50 & -3.61 & 12.11 & 1.70 & 0.102 & 0.57 & 0.87 & Anecdotal $H_0$ \\
~ & Right Hip Flexion (Planar) & 17 / 27 & 9.90 & -0.55 & 10.45 & 1.41 & 0.171 & 0.47 & 0.50 & Anecdotal $H_0$ \\
~ & Side Sway (Planar) & 17 / 27 & 5.13 & 0.08 & 5.05 & 1.43 & 0.170 & 0.54 & 0.76 & Anecdotal $H_0$ \\
\midrule
Minisquat & Left Knee Flexion (Planar) & 17 / 27 & 1.30 & 0.42 & 0.87 & 0.12 & 0.904 & 0.04 & 0.15 & Moderate $H_0$ \\
~ & Right Knee Flexion (Planar) & 17 / 27 & 0.64 & 3.24 & -2.61 & -0.32 & 0.749 & -0.10 & 0.16 & Moderate $H_0$ \\
~ & Left Hip Flexion (Planar) & 17 / 27 & 11.79 & -0.99 & 12.78 & 1.46 & 0.154 & 0.47 & 0.51 & Anecdotal $H_0$ \\
~ & Right Hip Flexion (Planar) & 17 / 27 & 10.82 & -0.99 & 11.82 & 1.36 & 0.184 & 0.43 & 0.42 & Anecdotal $H_0$ \\
~ & Lumbar Flexion (Planar) & 17 / 27 & 15.56 & 2.94 & 12.62 & 1.59 & 0.128 & 0.57 & 0.86 & Anecdotal $H_0$ \\
~ & Pelvis to Ankle Min Dist (cm) & 17 / 27 & -6.43 & -1.47 & -4.96 & -0.76 & 0.455 & -0.25 & 0.22 & Moderate $H_0$ \\
\bottomrule
\end{tabular}
}
\end{table}

\begin{table}[htbp]
\renewcommand\tablename{Table S7}
\renewcommand{\thetable}{\hspace{-.333333em}}
\centering
\caption{\textbf{Linear Mixed-Effects Model Results for 30-Day BeADS Monitoring.} The model evaluates the main effects of Group (Healthy vs. Patients) and Time (Day), alongside their interaction, while accounting for subject-level random intercepts.}
\label{tab:full_lme_beads}
\resizebox{\textwidth}{!}{
\begin{tabular}{llrrrr}
\toprule
\textbf{Metric} & \textbf{Term} & \textbf{Coefficient} & \textbf{Std. Error} & \textbf{z-value} & \textbf{p-value} \\
\midrule
Wrist to Ankle Min Dist (cm) & Intercept (Healthy, First Day) & 14.27 & 2.18 & 6.54 & $< 0.001$ \\
& Main Effect: Group (Patients) & 11.56 & 3.16 & 3.66 & $< 0.001$ \\
& Main Effect: Time (Day) & -0.12 & 0.02 & -5.93 & $< 0.001$ \\
& Interaction: Group $\times$ Time & 0.03 & 0.03 & 1.06 & 0.291 \\
\midrule
Lumbar Flexion Angle (Planar, in degrees) & Intercept (Healthy, First Day) & 109.17 & 2.44 & 44.67 & $< 0.001$ \\
& Main Effect: Group (Patients) & -13.41 & 3.54 & -3.79 & $< 0.001$ \\
& Main Effect: Time (Day) & 0.10 & 0.02 & 4.25 & $< 0.001$ \\
& Interaction: Group $\times$ Time & -0.05 & 0.03 & -1.48 & 0.140 \\
\bottomrule
\end{tabular}
}
\end{table}

\begin{figure}[htbp]
    \renewcommand\figurename{Figure S6}
    \renewcommand{\thefigure}{\hspace{-.333333em}}
    \centering
    \includegraphics[width=\textwidth]{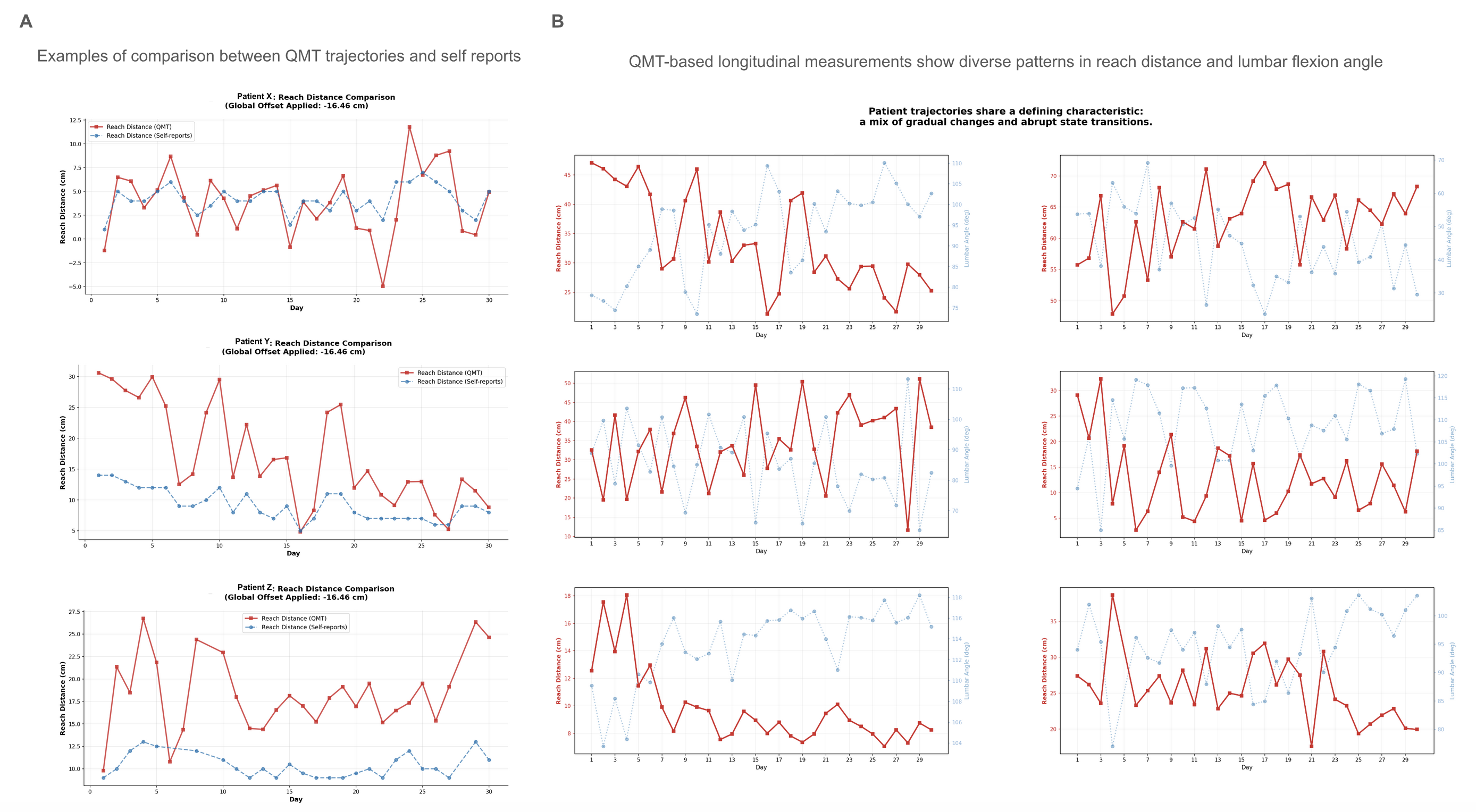} 
    \caption{\textbf{Longitudinal Reach Distance Trajectories.} (A) Individual subject comparisons between calibrated QMT estimates and subjective self-reports over the 30-day monitoring period. (B) QMT-derived reach distances and lumbar flexion angles over the 30 day period for different patients.}
    \label{fig:FigS6}
\end{figure}

\begin{figure}[htbp]
    \renewcommand\figurename{Figure S7}
    \renewcommand{\thefigure}{\hspace{-.333333em}}
    \centering
    \includegraphics[width=0.6\textwidth]{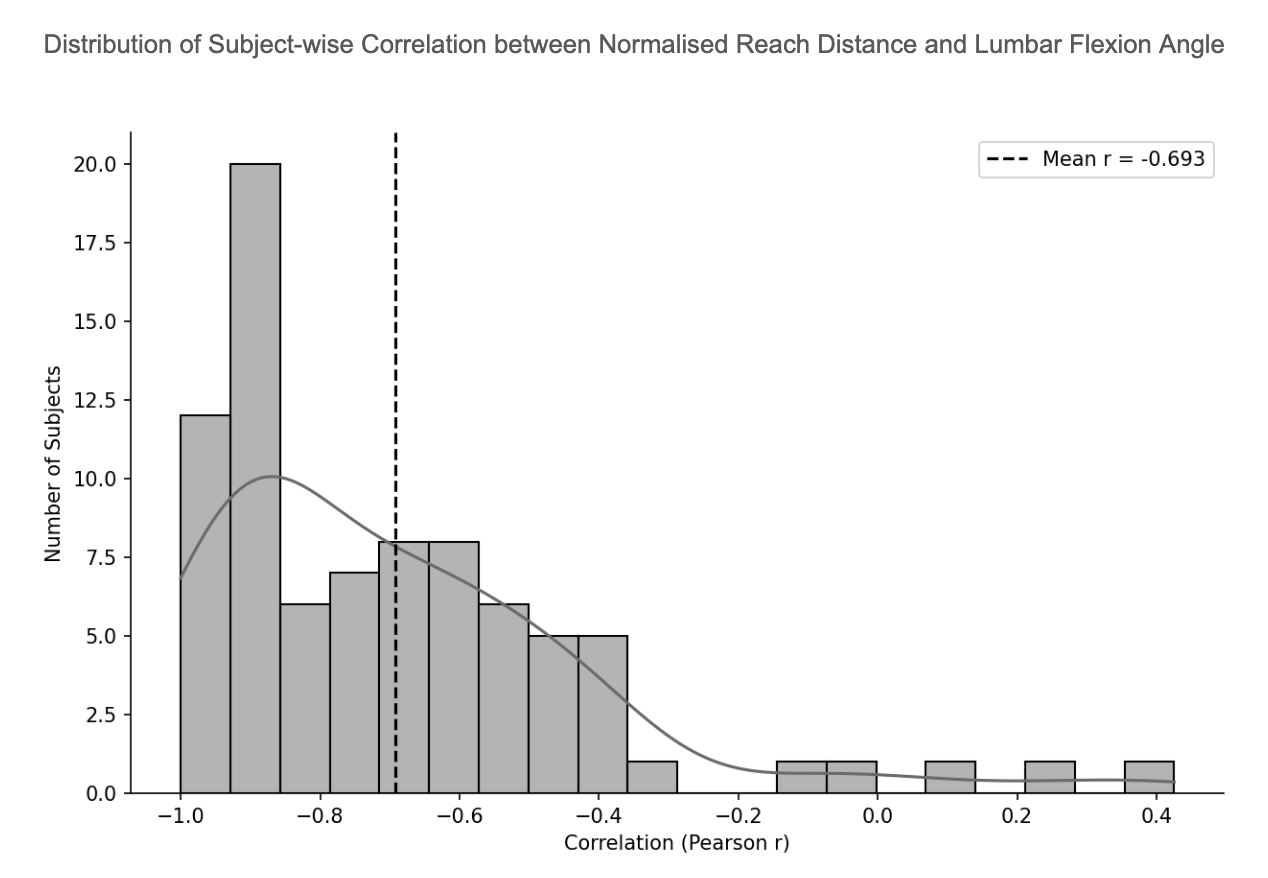} 
    \caption{\textbf{Distribution of Subject-wise Correlation between Normalised Reach Distance and Lumbar Flexion Angle}. Note that more flexion means lower Reach Distance and higher Lumbar Flexion Angle; therefore, they are inversely correlated. Both measurements are z-score normalised before computing the correlation.}
    \label{fig:FigS7}
\end{figure}

\begin{figure}[htbp]
    \renewcommand\figurename{Figure S8}
    \renewcommand{\thefigure}{\hspace{-.333333em}}
    \centering
    \includegraphics[width=0.6\textwidth]{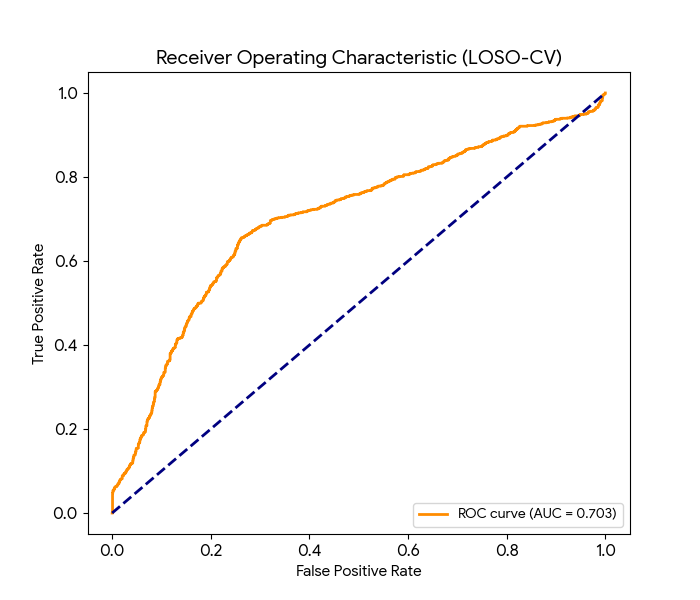} 
    \caption{\textbf{Receiver Operator Characteristics (ROC) Curve for classification between healthy controls and sciatica patients from BeADS dataset}. Leave one subject out cross-validation (LOSO-CV) accuracy is 68.52\%. Area under the ROC curve is 0.771. Classification uses a simple logistic regression model with following two QMT outputs: minimum wrist-to-ankle reach distance and maximum lumbar flexion angle, from a single day, as the inputs to the classifier. Low accuracy may be attributed to the difficulty in predicting patient class using just two metrics from a single day.}
    \label{fig:FigS8}
\end{figure}

\end{document}